\documentclass[margin,12pt]{article}
\usepackage[catalan,english]{babel}
\usepackage{lmodern}
\usepackage[T1]{fontenc}
\usepackage{latexsym}
\usepackage{amsfonts}
\usepackage{amsmath}
\usepackage{amssymb}
\usepackage[mathscr]{eucal}
\usepackage{a4wide}
\usepackage{graphicx}
\usepackage[usenames]{color}
\usepackage{slashbox}
\usepackage{xcolor}
\usepackage{bbold}
\usepackage{bbm}
\usepackage{authblk}
\usepackage{multirow}
\RequirePackage{hyperref}
\hypersetup{
	linktocpage,
    colorlinks,
    citecolor=blue,
    filecolor=black,
    linkcolor=blue,
    urlcolor=blue,
}

\title{\begin{flushright}
\end{flushright}
\vskip 0.1cm
{\bf{Study of $B\to\pi\ell\nu_{\ell}$ and $B^{+}\to\eta^{(\prime)}\ell^{+}\nu_{\ell}$ decays and determination of $|V_{ub}|$}}}
\author[a]{Sergi Gonz\`{a}lez-Sol\'{i}s\thanks{sgonzalez@itp.ac.cn}}
\author[b]{Pere Masjuan\thanks{masjuan@ifae.es}}
\affil[a]{{\it{CAS Key Laboratory of Theoretical Physics, Institute of Theoretical Physics, Chinese Academy of Sciences, Beijing 100190, China}}}
\affil[b]{{\it{Grup de F\'{i}sica Te\`{o}rica, Departament de F\'{i}sica, Universitat Aut\`{o}noma de Barcelona, Institut de F\'{\i}sica d'Altes Energies (IFAE), The Barcelona Institute of Science and Technology, Campus UAB, 08193 Bellaterra (Barcelona), Spain}}}

\def\be{\begin{equation}}
\def\ee{\end{equation}}
\def\bea{\begin{eqnarray}}
\def\eea{\end{eqnarray}}
\def\ben{\begin{enumerate}}
\def\een{\end{enumerate}}

\begin{document}
\maketitle

\abstract{We reassess the $B\to\pi\ell\nu_{\ell}$ differential branching ratio distribution experimental data released by the BaBar and Belle Collaborations supplemented with all lattice calculations of the $B\to\pi$ form factor shape available up to date obtained by the HPQCD, FNAL/MILC and RBC/UKQCD Collaborations.
Our study is based on the method of Pad\'{e} approximants, and includes a detailed scrutiny of each individual data set that allow us to obtain $|V_{ub}|=3.53(8)_{\rm{stat}}(6)_{\rm{syst}}\times10^{-3}$. 
The semileptonic $B^{+}\to\eta^{(\prime)}\ell^{+}\nu_{\ell}$ decays are also addressed and the $\eta$-$\eta^{\prime}$ mixing discussed.
}

\newpage

\section{Introduction}

Quark flavour-changing transitions in the Standard Model are described by the Cabibbo-Kobayashi-Maskawa (CKM) matrix whose elements, $V_{ij}$, weight the strength of the interaction. 
The CKM matrix satisfies unitarity imposing $\sum_{i}V_{ij}V_{ik}^{*}=\delta_{jk}$ and $\sum_{j}V_{ij}V_{kj}^{*}=\delta_{ik}$. 
To verify these relations a precise determination of the magnitude of the CKM elements becomes of capital importance since an eventual deviation of unitarity of the CKM matrix may be a hint of new physics. 
The most common (unitarity triangle) combination to look at is
\bea
V_{ud}V_{ub}^{*}+V_{cd}V_{cb}^{*}+V_{td}V_{tb}^{*}=0\, ,
\eea  
which contains the best-known side quantity $V_{cd}V_{cb}^{*}$ but also involves $V_{ub}$, one of the least-known elements.
The inclusive, $B\to X_{u}\ell\nu_{\ell}$, and exclusive, $B\to\pi\ell\nu_{\ell}$, semileptonic decays of a $B$ meson represent an advantageous laboratory to determine the value of $|V_{ub}|$ yielding the most precise value up to date.
Inclusive determinations are based, for example, on the Operator Product Expansion and perturbative QCD while exclusive determinations require knowledge of the shape of the participant meson Form Factor (FF) as a function of $q^2$, yielding the hadronic transition.
Numerically, the 2015 PDG reported values showed a $3.1\sigma$ deviation between the inclusive, $|V_{ub}|=(4.41\pm0.15^{+0.15}_{-0.17})\times10^{-3}$, and the exclusive, $|V_{ub}|=(3.28\pm0.29)\times10^{-3}$ \cite{Vubpdg2014}, determinations with a resulting average of $|V_{ub}|=(4.13\pm0.49)\times10^{-3}$.
The updated 2016 PDG version \cite{Agashe:2014kda} reports, respectively, $|V_{ub}|=(4.49\pm0.16^{+0.16}_{-0.18})\times10^{-3}$ and $|V_{ub}|=(3.72\pm0.19)\times10^{-3}$ \cite{Lattice:2015tia} for the inclusive and exclusive decays whose deviation, $2.6\sigma$, has slightly been reduced due to the one-$\sigma$ shift of the exclusive result.
At present, the PDG reports \cite{Olive:2016xmw}, respectively, $|V_{ub}|=(4.49\pm0.15^{\quad\,+0.16}_{\rm{exp}-0.17_{th}}\pm0.17)\times10^{-3}$ and $|V_{ub}|=(3.70\pm0.10\pm0.12)\times10^{-3}$ \cite{Amhis:2016xyh} for the inclusive and exclusive determinations.
The origin of this long-standing discrepancy between the inclusive and exclusive determinations still remains unclear, demanding the resulting combined average, $|V_{ub}|=(3.94\pm0.36)\times10^{-3}$ \cite{Olive:2016xmw}, to be borrowed with caution.
As pointed out already in Ref.\,\cite{Neubert:2008cp}, and recently adopted in Ref.~\cite{Crivellin:2014zpa}, a new physics explanation of this tension is very unlikely and, therefore, it might be due to an underestimation of uncertainties in the experimental and/or theoretical analysis. 

In this work, we reexamine the exclusive $B\to\pi\ell\nu_{\ell}$ decays and extract $|V_{ub}|$ following an alternative approach, regarding the parameterization of the participant vector form factor, slightly different than the most commonly used $z$-expansion and Vector Meson Dominance models, and profiting from the large set of experimental data and lattice simulations.
A detailed scrutiny of each individual data set, explored bin by bin, allows us to identify agreements and tensions among them and propose a path towards further determinations.

The hadronic matrix current for the $B \to \pi \ell \nu_{\ell}$ decay can be written as
\begin{equation}
\langle\pi(p_{\pi})|V_{\mu}|B({p_{B}})\rangle=f_{+}(q^{2})\left(p_{B}+p_{\pi}-q\frac{m_{B}^{2}-m_{\pi}^{2}}{q^{2}}\right)_{\mu}+f_{0}(q^{2})\frac{m_{B}^{2}-m_{\pi}^{2}}{q^{2}}q_{\mu}\,,
\end{equation}
where $q=p_{B}-p_{\pi}=p_{\ell}+p_{\nu_{\ell}}$ is the transferred momentum to the dilepton pair while $f_{+}(q^{2})$ and $f_{0}(q^{2})$ are, respectively, the participant vector and scalar form factors encoding the dynamics of the strong interactions occurring in the heavy-to-light $B\to\pi$ hadronic transition.

For light leptons ($e$ and $\mu$), one can safely take the $m_{\ell}\to0$ limit so that only the $f_{+}(q^{2})$ is relevant and the corresponding partial decay width distribution is given by\footnote{The interested reader is referred to Ref.\,\cite{Yao:2018tqn} for a recent parameterization of the scalar form factor $f_{0}(q^{2})$ based on dispersion relations.}
\bea
\frac{d\Gamma(B\to\pi\ell\nu_{\ell})}{dq^{2}}=\frac{G_{F}^{2}|V_{ub}|^{2}}{192\pi^{3}m_{B}^{3}}|p_{\pi}|^{3}|f_{+}(q^{2})|^{2}\,,
\label{decayrate}
\eea
where $p_{\pi}=\sqrt{(m_{B}^{2}+m_{\pi}^{2}-q^{2})^{2}-4m_{B}^{2}m_{\pi}^{2}}$ is a kinematical factor accounting for the momentum of the pion in the $B$ meson rest frame.
The main source of uncertainty in the extraction of $|V_{ub}|$ lies on the vector form factor which, in turn, requires a reliable parameterization in terms of $q^2$. 


From the experimental side, the CLEO Collaboration reported the first measurement of the $B\to\pi\ell\nu_{\ell}$ branching fraction in 1996~\cite{Alexander:1996qu}, later updated in 2003~\cite{Athar:2003yg}, and released the partial branching ratio distribution measured in 4 $q^2$ bins in 2007~\cite{Adam:2007pv}. 
More recently, the $q^{2}$ decay spectra have been measured, respectively, in 6-and 12-bins of $q^{2}$ by BaBar in 2011~\cite{delAmoSanchez:2010af} and 2012~\cite{Lees:2012vv}, and by Belle in 13-bins in 2011 \cite{Ha:2010rf} and in 13-and 7-bins for the $B^{0}$ and $B^{-}$ mode, respectively, in 2013~\cite{Sibidanov:2013rkk}.

On the lattice QCD side, results on the form factor shape at large $q^{2}$ were obtained by the HPQCD Coll. in 2007~\cite{Dalgic:2006dt} and by the FNAL/MILC Coll. in 2008~\cite{Bailey:2008wp} in 5-and 12-bins of $q^{2}$, respectively. 
In 2015 the RBC/UKQCD Coll. released new results in 3 bins of $q^{2}$~\cite{Flynn:2015mha} and the FNAL/MILC Coll. presented an updated analysis~\cite{Lattice:2015tia}.

In total, we have a set of five experimental measurements of the $B\to\pi\ell\nu_{\ell}$ decay spectra driving the form factor shape at small $q^{2}$ and a set of four lattice QCD simulations for the form factor dominating the large $q^{2}$ region.
In order to determine $|V_{ub}|$ with good precision (beyond $10\%$), it is desirable to have a suitable parameterization of the intermediate energy region $(15$-$20$ GeV$^{2})$ connecting both the small-and large-$q^{2}$ regions in a continuous and derivable way, under the constraints of unitary and analyticity. 

From a theoretical perspective, parameterizations based on resonance-exchange ideas~\cite{Wirbel:1985ji,Grinstein:1986ad,Nussinov:1986hw,Suzuki:1987ap} have been widely used so far to describe the $B\to\pi$ FF shape.
The parameterizations proposed by Be\'{c}irevic-Kaidalov~\cite{Becirevic:1999kt} and Ball-Zwicky~\cite{Ball:2004ye}, incorporating some properties of the FF such the value of the kinematical constraint at $q^{2}=0$ and the position of the $B^{*}$ pole in the old-times spirit~\cite{Wirbel:1985ji,Grinstein:1986ad,Nussinov:1986hw,Suzuki:1987ap}, became rather popular in the first decennial of this century.
Both descriptions contain free parameters, such additional poles that pick up effects of multi-particle states, to be fixed from fits to experimental data. 
However, the election of these ans\"{a}tze induces a source of theoretical (or systematic) uncertainty difficult to quantify. 
Moreover, as argued in Ref.~\cite{Bailey:2008wp}, if the reconstruction of the FF obtained only from fits to experimental data is seen inconsistent with the shape derived by the lattice Collaborations, one would not unveil whether experiment and theory disagree or simple parameterizations are insufficient.
To improve on that, the so-called $z$-parameterization was proposed~\cite{Okubo:1971jf,Boyd:1994tt,Bourrely:2008za}. 
This is based on a conformal transformation expansion which guarantees unitary constraints on its coefficients,  even though in practice the constraints are rather weak.


Lets us return to the 2015 and 2016 PDG editions reported values for $|V_{ub}|$ from exclusive processes, $(3.28\pm0.29)\times10^{-3}$ and $(3.72\pm0.19)\times10^{-3}$, respectively.
They were obtained from simultaneous fits to the four most precise measurements of BaBar~\cite{delAmoSanchez:2010af,Lees:2012vv} and Belle~\cite{Ha:2010rf,Sibidanov:2013rkk} together with the 2008 and 2015 MILC Collaboration lattice simulations on the FF, respectively. 
While the 2015 PDG value corresponded to the determination provided by the HFAG as of summer 2014~\cite{Vubpdg2014}, the updated 2016 PDG version reports the value obtained by the MILC Collaboration in 2015~\cite{Lattice:2015tia}.
These two values have been determined by using a $z$-parameterization as fit function and show a sizable deviation of $1.3~\sigma$ whose origin stems mainly from the following fact.
While the lattice FF simulations from the MILC Collaboration in 2008 \cite{Bailey:2008wp} were included into the fit in the HFAG analysis of 2014, and on top of that only 4 of the 12 points were used to avoid correlations between neighboring points, the result obtained by MILC in 2015 considers their updated FF simulation ones \cite{Lattice:2015tia}.
Moreover, while the 7 bins of the $B^{-}$ decay mode measurement reported by Belle in 2013 were not included into the HFAG 2014 fit, the MILC 2015 include them into their analysis.

Previous PDG reported values, e.g. $|V_{ub}|=(3.23\pm0.30)\times10^{-3}$ in PDG 2012, showed the corresponding HFAG fit results obtained from simultaneous fits to the existent experimental measurements at the time together with the MILC form factor predictions of 2008 using 6 of the 12 points instead of 4 of 12 as in the HFAG result of 2014.
Both the choice of the MILC 2008 number-and bin-points to fit and the omission of the HPQCD form factor lattice simulations of 2007 is not rather clear to us.
Accepting the FNAL/MILC lattice form factor calculation of 2015 presents several improvements with respect to their 2008 predictions,\footnote{The 2015 updated analysis shifted downward the central values of the FF and reduced the error by almost a factor of three with respect to the study of 2008.} still the theoretical error associated to the FF represents the largest uncertainty in $|V_{ub}|$.
In this respect, the lattice simulation provided by the RBC/UKQCD Collaboration~\cite{Flynn:2015mha} has been welcomed, obtaining $|V_{ub}|=(3.61\pm0.32_{\rm{stat+syst}})\times10^{-3}$ from a combined fit to their results for the form factor together with BaBar and Belle experimental data.

More recently, Ref.~\cite{Dingfelder:2016twb} obtained $|V_{ub}|=(3.59\pm0.12_{\rm{stat}})\times10^{-3}$ from a combined fit to BaBar and Belle data and the FNAL/MILC form factor simulations of 2015 supplemented by the light-cone sum rule (LCSR) result at $q^{2}=0$ GeV$^{2}$ \cite{Bharucha:2012wy}, while the HFAG update for the summer 2016 consists in employing an average $q^{2}$ in experimental and lattice data (including this time the measured $q^{2}$ spectra of the $B^{-}$ decay mode of Belle 2013) with the LCSR prediction at $q^{2}=0$ GeV$^{2}$ \cite{Bharucha:2012wy} obtaining $|V_{ub}|=(3.65\pm0.09_{\rm{exp}}\pm0.11_{\rm{th}})\times10^{-3}$ \cite{HFAG2016}.

In 2016, the FLAG working group reported $|V_{ub}|=(3.62\pm0.14)\times10^{-3}$ from a fit to lattice and BaBar and Belle experimental data~\cite{Aoki:2016frl} while the current PDG edition reports the HFLAV value of 2017 $|V_{ub}|=(3.70\pm0.10\pm0.12)\times10^{-3}$ \cite{Olive:2016xmw,Amhis:2016xyh} obtained from an averaged $q^{2}$ spectrum of all BaBar and Belle data sets constraining the $\chi^{2}$ minimization by averaged values for the coefficients of the form factor parameterization derived by the lattice groups and by the LCSR prediction at $q^{2}=0$.

Finally, a closer look to the plots of the corresponding fit results of the different analyses reveals a discrepancy between HFAG 2014 and Ref.~\cite{Dingfelder:2016twb}, and lattice groups~\cite{Lattice:2015tia,Flynn:2015mha} on the position of the last experimental datum of both BaBar 2011 and BaBar 2012 measurements\footnote{This point has the smaller error out of the 6 (12) BaBar 2011(12) data points. 
The $q^{2}$ average of experimental data of HFAG 2016, however, place this point in the middle of each bin as in Refs.~\cite{Lattice:2015tia,Flynn:2015mha}.}.

Although the different $|V_{ub}|$ determinations are consistent with each other, we find the situation slightly unclear and without consensus among different groups regarding the use of experimental and theoretical data to fit. 

The main purpose of this work is to reanalyze the $B\to\pi\ell\nu_{\ell}$ experimental data and discuss the impact of including into the fit each of the lattice-QCD simulations on the FF shape.
We use the method of Pad\'{e} approximants (PAs in what follows) to parameterize the $B\to\pi$ transition.
These provide for a model-independent method, simple and user-friendly, with the important advantage of incorporating FF's unitary and analyticity constraints by construction, thus providing a systematic error. 

We have discussed the Pad\'{e} method in Refs.~\cite{Masjuan:2007ay,Masjuan:2008fv,Masjuan:2012wy} and illustrated its usefulness as fitting functions in Refs.\,\cite{Escribano:2013kba,Escribano:2015nra,Escribano:2015yup,Escribano:2015vjz} applied to the description of the $\pi^{0},\eta$ and $\eta^{\prime}$ transition form factors. 
In these cases, the approximants showed an interesting ability to connect the low-and high-energy realms while improving the description of part of the intermediate-energy regime.
The method allows us here to obtain a value for $|V_{ub}|$, including both statistical and systematic uncertainties coming from the fit function, with a stamp of model independence.
Constraints from unitary of the form factor will show up naturally and will provide for a roadmap towards next steps to follow both for theoretical as well experimental studies.

Although being the most precise, $B\to\pi\ell\nu_{\ell}$ only amounts to $\sim7\%$ of the $B\to X_{u}\ell\nu_{\ell}$ decays. 
Measurements of the branching fractions distributions of $B^{+}\to\omega\ell^{+}\nu_{\ell}$ and of $B^{+}\to\eta\ell^{+}\nu_{\ell}$ in $5$ bins of $q^{2}$ were released in 2012, and the branching ratio of $B^{+}\to\eta^{\prime}\ell^{+}\nu_{\ell}$ reported, by the semileptonic charmless program of BaBar \cite{Lees:2012vv}.
In the second part of this work, we will tackle the $B^{+}\to\eta^{(\prime)}\ell^{+}\nu_{\ell}$ decays, predicting the differential branching ratio distributions and extracting the $\eta$-$\eta^{\prime}$ mixing angle taking advantage of the $B\to\pi$ form factor parameterizations obtained in the first part of this work.

As a final introductory remark, we shall mention that a method based on dispersion theory to extract $|V_{ub}|$ from the $B_{\ell4}$ decay has been proposed in Ref.\,\cite{Kang:2013jaa}.

This article is structured then as follows: In section \ref{formfactor} we address the analytical structure of the participant $B\to\pi$ form factor and discuss the most common theoretical descriptions that have been considered in literature so far. 
In this section we also present our proposal, a parameterization based on the unitary and analyticity of the FF which allows us to use a sequence of PAs. 
In section \ref{section1}, we show our fit results to the BaBar, Belle and CLEO differential branching ratio distribution experimental data which enables us to determine the product $|V_{ub}f_{+}(0)|$ and extract, subsequently, $|V_{ub}|$ by using the LCSR prediction of $f_{+}(0)$ given in Ref.\,\cite{Bharucha:2012wy}. 
In section \ref{section4} we discuss the impact of including the different lattice QCD predictions on the FF shape into the analysis and determine $|V_{ub}|$ directly from a simultaneous fit. 
In this section we present our central fit results, evaluate the role of introducing the value of $f_{+}(0)$ as an additional restriction in the $\chi^{2}$ minimization, and perform fits to the lattice data alone.
Unitary constraints on the Pad\'{e} approximants are discussed in section \ref{unitaryconstraints}.
In section \ref{section5} we predict the $B^{+}\to\eta^{(\prime)}\ell^{+}\nu_{\ell}$ differential branching fractions distributions and determine the $\eta$-$\eta^{\prime}$ mixing.
Finally, our conclusions are devoted to section \ref{conclusions}.

Preliminary results of this study have been presented in Refs.\,\cite{Gonzalez-Solis:2016vea,Gonzalez-Solis:2018dyl}.

\section{$B\to\pi$ form factor}\label{formfactor}

A form factor is an analytic function everywhere in the complex plane except for isolated poles and branch cuts. 
Poles correspond to single particle intermediate states while branch cuts originate when the energy reaches a threshold for producing multi-particle intermediate states. 
For the $B\to\pi\ell\nu_{\ell}$ decay concerning us, the lightest production 
threshold is located at $s_{\rm{th}}=(m_{B}+m_{\pi})^{2}$ GeV$^{2}$, lying slightly above the available kinematical energy range of the decay, $0<q^{2}<(m_{B}-m_{\pi})^{2}$ GeV$^{2}$.
A first approximation to the form factor suggests a single pole description driven by the exchange of a $\bar{u}b$ intermediate state, the $B^{*}$ meson with mass $m_{B^{*}}=5.325$ GeV (with very small width) and quantum numbers $J^{P}=1^{-}$.
For illustrative purposes, let us consider a dispersive representation of the form factor in terms of $q^{2}$, where $q^{2}$ is the invariant mass of the lepton pair,
\bea
f_{+}(q^{2})=\frac{1}{\pi}\int_{s_{\rm{th}}}^{\infty}ds^{\prime}\frac{{\rm{Im}}f_{+}(s^{\prime})}{s^{\prime}-q^{2}-i\varepsilon}\,,
\label{dispersionrelation}
\eea
with Im$f_{+}(s) = \pi \rho(s)$. 
The single pole description in Eq.~(\ref{dispersionrelation}) would correspond to using for the spectral function $\rho(s)=f_{+}(0)m_{B^*}^2 \delta(s-m_{B^{*}}^2)$.
This give raise to the Vector Meson Dominance model (VMD) with a $B^{*}$ pole appearing between the available phase space and the lowest production threshold, $(m_{B}-m_{\pi})^{2}<s_{p}<(m_{B}+m_{\pi})^{2}$ GeV$^{2}$,
\begin{equation}
f_{+}(q^{2})=\frac{f_{+}(0)}{1-q^2/m_{B^{*}}^2}\,,
\label{VMD}
\end{equation}
where $f_{+}(0)$ is a normalization constant.


However, this model obviates effects of heavier vector states. Be\'{c}irevi\'{c} and Kaidalov (BK) \cite{Becirevic:1999kt} proposed a modification of the VMD via including, above $q^{2}=(m_{B}+m_{\pi})^{2}$ GeV$^{2}$, a heavier narrow-width resonance, a $B^{*\prime}$, through adding ${\rm{Im}}f_{+}(s)=\pi\rho(s)\propto\delta(s-m_{B^{*\prime}}^{2})$ in Eq.~(\ref{dispersionrelation}), leading to
\bea
f_{+}(q^{2})=\frac{r_{1}}{1-q^{2}/m_{B^{*}}^{2}}+\frac{r_{2}}{1-q^{2}/m_{B^{*\prime}}^{2}}\,.
\eea
Implementing that the form factor behaves as $1/q^4$ at large $q^2$ together with $f_{+}(0) = r_1 + r_2$, 
the standard expression for the BK form factor reads
\bea
f_{+}(q^{2})=\frac{f_{+}(0)}{(1-q^{2}/m_{B^{*}}^{2})(1-\alpha \, q^{2}/m_{B^{*}}^{2})}\,,
\label{BKFF}
\eea
where $\alpha$ fixes the position of the second fitted \textit{effective} pole.

Later on, Ball and Zwicky (BZ) \cite{Ball:2004ye,Ball:2005tb} proposed a similar expression in terms of three parameters $\{f_{+}(0),r,\alpha\}$ by imposing the form factor to fall-off as $\sim1/q^2$ at large $q^2$ instead.
The matching $f_{+}(0)=r_{1}+r_{2}$ and $r=r_2(\alpha-1)$ leads
\bea
f_{+}(q^{2})=\frac{f_{+}(0)}{1-q^{2}/m_{B^{*}}^{2}}+\frac{r \,q^{2}/m_{B^{*}}^{2}}{(1-q^{2}/m_{B^{*}}^{2})(1-\alpha \, q^{2}/m_{B^{*}}^{2})}\,,
\label{BZFF}
\eea
where $r$ may be understood as a parameter which encodes the relative weight of the second \textit{effective} resonance with respect to the first one.

The above two parameterizations fix the position of the $B^{*}$ pole to its mass, $m_{B^{*}}=5.325$ GeV, while the rest of free parameters, $\{f_{+}(0),\alpha\}$ and $\{f_{+}(0),r,\alpha\}$ in Eqs.~(\ref{BKFF}) and (\ref{BZFF}), respectively, are inferred from fits to experimental data.


Exploiting the analyticity and positivity properties of the vacuum polarization functions, Okubo and collaborators proposed the method of unitary bounds~\cite{Okubo:1971jf} in the context of kaon decays, which later on was applied for semileptonic B decays~\cite{Bourrely:2008za,Bourrely:1980gp}.
This method, called $z$-parameterization and reviewed in Refs.~\cite{Boyd:1994tt,Arnesen:2005ez}, parameterizes $f_{+}(q^{2})$ as a Taylor expansion in terms of a conformal complex variable $z$ as follows:
\begin{eqnarray}\label{zparam}
f_{+}(q^2)&=&\frac{1}{P(q^2) \phi(q^2,q_0^2)} \sum_{n=0}^{\infty} a_n(q_0^2)[z(q^2,q_0^2)]^n\,,
\end{eqnarray}
where
\begin{eqnarray}
z(q^2,q_0^2)&=&\frac{\sqrt{1 - q^2/t_+}-\sqrt{1 - q_0^2/t_+}}{\sqrt{1 - q^2/t_+}+\sqrt{1 - q_0^2/t_+}}\,,
\end{eqnarray}
with $t_+ = (m_B + m_\pi)^2$ GeV$^{2}$ and $\phi(q^2,q_0^2)$ a function given in Ref.~\cite{Boyd:1994tt}. 
The function $P(q^2)=z(q^2,m_{B*}^2)$ is the Blaschke factor which accounts for the pole at $q^2=m_{B*}^2$. 
The free parameter $q_0^2$ is chosen to optimize the fit. 
Assuming the spectral function driving the FF to be saturated by $B\pi$ vector intermediate states, unitary and crossing symmetry guarantee the coefficients $a_n(q_0^2)$ to satisfy $\sum_{n=0}^{\infty} a_n^2(q_0^2) \leq 1$. 

In practice, Eq.~(\ref{zparam}) is truncated at a finite order (typically up to first or second order) which implies the FF to behave as $\sim 1/q^4$ at large $|q^2|$ due to $\phi(q^2,q^2_0)$, in contradiction with perturbative QCD scaling~\cite{Nussinov:1986hw,Suzuki:1987ap}. Beyond, as discussed in \cite{Bourrely:2008za}, the outer function has an unphysical singularity at the $B\pi$ production threshold $t_+$. 
This unphysical singularity may distort the behavior near the upper end of the physical region, where the FF is poorly known. 
These considerations triggered an alternative $z$-parameterization proposed in~\cite{Bourrely:2008za} by Bourrely-Caprini-Lellouch (BCL):
\begin{eqnarray}\label{zparam2}
f_{+}(q^2) = \frac{1}{1-q^2/m_{B*}^2} \sum_{n=0}^{N-1} b_{+}^{(n)}\left[z(q^2,q_0^2)^{n}-(-1)^{n-N}\frac{n}{N}z(q^2,q_0^2)^{N}\right]\, ,
\end{eqnarray}
where the pole included by hand ensures the correct analytic structure in the complex plane and the proper scaling, $f_{+}(q^2) \sim 1/q^2$ at large $q^2$. 

Let us comment that the $z$-parameterization is not a zero-preserving transformation with respect of $q^2$ unless the particular choice $q_0^2=0$ is made, which implies $z \to 0$ does not come from $q^2 \to 0$, but rather from a large $q^2$ value. 
This poses a word of caution when using the $z$-parameterization to determine the behavior of the FF at low $q^2$. 
We shall add here that the definition of $z(q^2,q_0^2)$ corresponds formally with a Quadratic approximant, a well-defined extension of a PA that includes square-root terms~\cite{Gonzalez-Solis:2016vea,QA}\footnote{In particular, $z(q^2,0) ={\cal Q}_{1,1}^1(q^2)$ with coefficients $R_0 =-1$, $S_0 = Q_0=0$, $R_1=S_1=Q_1=1/2$~\cite{Gonzalez-Solis:2016vea}.}. 
As such, it is formally a PA of order given by the truncated series, either in Eq.~(\ref{zparam}) or (\ref{zparam2}), for which the PA convergence constraints must be applied.





To complete the overview, the AFHNV approach~\cite{Flynn:2007qd,Flynn:2007ii} is based on the Omn\`es representation which expresses the analytic function in terms of its phase along the boundary of the analyticity domain. 
If one takes into account the pole $q^2=m_{B*}^2$, assumes that the FF have no zeros in the complex plane, and by Watson's theorem that the phase $\delta(t)$ is equal, below the first inelastic threshold, to the phase of the $P$- wave with $I=1/2$ of the $\pi B \to \pi B$ elastic scattering, the representation reads (assuming $n\gg1$ which implies a multiply-subtracted dispersion relation-and neglects altogether the dispersive integral):
\begin{equation}
f_{+}(q^2) =\frac{1}{m_{B*}^2-q^2} \prod_{j=1}^n [f_{+}(q^2_j)(m_{B*}^2-q_j^2)]^{\alpha_j(q^2)}\,, 
\end{equation}
\noindent
where $\alpha_j(q^2) = \prod_{i=0,i\neq j}\frac{q^2-q^2_i}{q^2_j-q^2_i}$. Notice that after the multiply-subtracted dispersion relation the exponential behavior at large $q^2$ does not corresponds to the one from QCD and, due to the lack of the dispersive integral, the original branch cut $q^2\leq t_+$. 


\subsubsection*{Our proposal: Pad\'{e} approximants}

The form factor $f_{+}(q^2)$ is a Stieltjes function, which is a function that can be represented by an integral form defined as~\cite{Baker}
\begin{equation}\label{Stieltjes}
f_{+}(q^2) = \int_0^{1/R} \frac{\rm{d} \phi(u)}{1-u q^2}\,,
\end{equation}
where $\phi(u)$ is any bounded and non-decreasing function. 
By defining $R=(m_B+m_{\pi})^2$ GeV$^{2}$, identifying 
${\rm{d}}\phi(u) = \frac{1}{\pi}\frac{{\rm{Im}}f_{+}(1/u)}{u} \rm{d} u$, and making the change of variables $u=1/s$, Eq.~(\ref{Stieltjes}) returns the dispersive representation of the form factor given in Eq.~(\ref{dispersionrelation}). 
To be strict, Eq.~(\ref{dispersionrelation}) is a meromorphic function of Stieltjes type. 

Since the FF, and its imaginary part, is created by the vector current, Im$f_{+}(s)$  is a positive function, the requirement of $\phi(u)$ to be non-decreasing is fulfilled and the convergence of PAs to the FF is guaranteed. 

Pad\'e Theory not only provides a convergence theorem for a sequence of PAs to Stieltjes (or Stieltjes-type) functions, i.e., $\lim_{N,M\to \infty} P^N_M(q^2) - f_{+}(q^2) =0$, but also its rate of convergence~\cite{Baker,Masjuan:2009wy}, given by the difference of two consecutive elements in the PA sequence. 
As we will see later, this error prescription will return very small theoretical uncertainties. 
To be more conservative, in Refs.~\cite{Masjuan:2008fv,Masjuan:2012wy,Escribano:2013kba,Escribano:2015nra}  we designed a different method to extract such uncertainty which yields errors at the level of the statistical ones.

Pad\'{e} approximants to a given function are ratios of two polynomials (with degree $M$ and $N$, respectively)\footnote{With any loss of generality, we take $b_{0}=1$ for definiteness.},
\begin{eqnarray}
P^{M}_{N}(q^{2})=\frac{\sum_{j=0}^{M}a_{j}(q^{2})^{j}}{\sum_{k=0}^{N}b_{k}(q^{2})^{k}}=
\frac{a_{0}+a_{1}q^{2}+\cdots+a_{M}(q^{2})^{M}}{1+b_{1}q^{2}+\cdots+b_{M}(q^{2})^{N}}\,,
\label{Pade}
\end{eqnarray}
with coefficients determined after imposing a set of a accuracy-through-order conditions with the function one wants to approximate, which is 
 $f_{+}(q^2)-P^{M}_{N}(q^{2})={\mathcal O}(q^2)^{M+N+1}$.
 
We would like to remark that Eqs.~(\ref{VMD}), (\ref{BKFF}), (\ref{BZFF}), and (\ref{zparam}) can be seen as particular elements of the general sequence of PAs given in Eq.~(\ref{Pade}).


Besides ordinary sequences of PAs we will also consider Pad\'{e} Type approximants $T_{N}^{M}(q^2)$ and Partial Pad\'{e} approximants $P_{Q,N-Q}^{M}(q^2)$ in our study.
 $T_{N}^{M}(q^2)$ have the denominator fixed in advanced (by imposing the location of the zeros of it), while $P_{Q,N-Q}^{M}(q^2)$ only $Q$ zeros of the denominator are fixed in advanced while the rest are left free. Strictly speaking then VMD, BK and BZ correspond, respectively, to the $T^{0}_{1}(q^{2})$, $P^{0}_{1,1}(q^{2})$ and $P^{1}_{1,1}(q^{2})$ elements, while the $z$-parameterization corresponds to a polynomial in terms of a ${\cal Q}_{1,1}^1$ as we argued before.
The main advantage of fixing a pole is that the number of parameters to fit decreases by one and typically allows to reach higher elements of the sequence~\cite{Masjuan:2008fv}. If the sequence is large enough and the position of the first singularity is accurately known, the convergence of the $T_{N}^{M}(q^2)$ is faster than the convergence of ordinary PAs for Stieltjes functions~\cite{Masjuan:2007ay,Baker,Masjuan:2008fr}. 



\section{Fits to experimental and lattice data}

\subsection{Fits to the $B\to\pi\ell\nu_{\ell}$ BaBar and Belle data}\label{section1}

Our first analysis consists of fitting the most recent $B\to\pi\ell\nu_{\ell}$ branching ratio distribution experimental data released by BaBar in 2011~\cite{delAmoSanchez:2010af} and 2012~\cite{Lees:2012vv} and by Belle in 2011~\cite{Ha:2010rf} and 2013~\cite{Sibidanov:2013rkk}.
We will also briefly discuss the effect of including CLEO 2007 results~\cite{Adam:2007pv} into the fit, which are usually neglected. 
In order to facilitate the reproduction of our results, we would like to write down from which Tables of the papers the experimental data we use come from.
For CLEO 2007 we use the results reported in Table I of Ref.~\cite{Adam:2007pv}\footnote{CLEO 2007 result consists in measurements of partial branching fractions in only 4 unequal $q^{2}$ subregions and no bin-to-bin correlation matrix is reported. 
For our analysis, we have placed each experimental datum at the middle of each of the corresponding subregions and scaled the bin values accordingly. 
}. 
For BaBar 2011 we employ the data collected in Tables X and XXVIII of Ref.~\cite{delAmoSanchez:2010af} while for BaBar 2012 we use Tables XXIII, XXVIII and XXXI of Ref.~\cite{Lees:2012vv}\footnote{For BaBar 2012, we use the combined analysis of both $B^{0}$ and $B^{-}$ modes assuming isospin symmetry.}. 
Regarding Belle 2011, we employ the data gathered in Tables III, IV and V of Ref.~\cite{Ha:2010rf}, respectively. 
Finally, for Belle 2013 we employ the data given, respectively, in Tables XVII, XVIII, XIX and XX of Ref.~\cite{Sibidanov:2013rkk}. 
To this later data we have added, as suggested in Table XII of \cite{Sibidanov:2013rkk}, a systematic uncertainty of $5.0\%$ and $5.1\%$ of the $q^{2}$ bin value for the $B^{+}$ and $B^{0}$ mode, respectively, and assumed a systematic correlation of the $49\%$ between the two modes as written in the paper below that Table.
For our study, we assume, for convenience, isospin symmetry to translate the Belle 2013 data on the $B^{-}$ mode to the $B^{0}$ ones through
\bea
\Delta(B^{0}\to\pi^{+}\ell^{-}\nu_{\ell})=2\frac{\tau_{B^{0}}}{\tau_{B^{-}}}\Delta(B^{-}\to\pi^{0}\ell^{-}\nu_{\ell})\,,
\eea 
where $\tau_{B^{0}}=(1.520\pm0.004)\times10^{-12}$ s and $\tau_{B^{-}}=(1.638\pm0.004)\times10^{-12}$ s, are, respectively, the mean life time of the neutral and charged $B$ mesons \cite{Agashe:2014kda}.
In all, we will treat the five experimental data sets as independent measurements, i.e., no statistical either systematic correlations between the five different analyses is considered\,\cite{Lattice:2015tia,Flynn:2015mha}.

The $\chi^{2}$ function minimized in our first fit is defined as
\begin{eqnarray}
\chi^{2}_{\rm{data}}&=&\sum_{i,j=1}^{6}\Delta_{i}^{{\rm{BaBar11}}}({\rm{Cov}}_{ij}^{\rm{BaBar11}})^{-1}\Delta_{j}^{\rm{BaBar11}}+\sum_{i,j=1}^{12}\Delta_{i}^{\rm{BaBar12}}({\rm{Cov}}_{ij}^{\rm{BaBar12}})^{-1}\Delta_{j}^{\rm{BaBar12}}\nonumber\\[1ex]
&+&\sum_{i,j=1}^{13}\Delta_{i}^{{\rm{Belle11}}}({\rm{Cov}}_{ij}^{\rm{Belle11}})^{-1}\Delta_{j}^{\rm{Belle11}}+\sum_{i,j=1}^{20}\Delta_{i}^{\rm{Belle13}}({\rm{Cov}}_{ij}^{\rm{Belle13}})^{-1}\Delta_{j}^{\rm{Belle13}}\,,\nonumber\\
\label{chi2data}
\end{eqnarray}
where
\bea
\Delta_{k}^{\rm{data}}=\left(\frac{\Delta B}{\Delta q^{2}}\right)^{\rm{data}}-\frac{1}{\Gamma_{B}}\frac{G_{F}^{2}}{192\pi^{3}m_{B}^{3}}|V_{ub}f_{+}(0)|^{2}|p_{\pi}|^{3/2}|\widetilde{P}_{N}^{M}(q^{2})|^{2}\, ,
\label{delta}
\eea
with $\Gamma_{B}$ the full width of the $B$ meson, ${\rm{Cov}}_{ij}$ denotes the corresponding covariance matrix and $\widetilde{P}_{N}^{M}(q^{2})=P_{N}^{M}(q^{2})/P_{N}^{M}(0)$ the PA normalized to unity at the origin of energies whose coefficients will be determined by the fit.

Notice that we do not access $|V_{ub}|$ but rather the product $|V_{ub}f_{+}(0)|$ that factors out in Eq.~(\ref{delta}).
However, we can still extract $|V_{ub}|$ by invoking external information on the $f_{+}(0)$.
In this work, we use $f_{+}(0)=0.261^{+0.020}_{-0.023}$ determined in a light-cone sum rules calculation \cite{Bharucha:2012wy}\footnote{Actually, for this quantity we use $f_{+}(0)=0.2595\pm0.0215$, where we have symmetrised the errors.}.
Other possible choices would be $f_{+}(0)=0.26^{+0.04}_{-0.03}$ \cite{Duplancic:2008ix}, $f_{+}(0)=0.281\pm0.033$ \cite{Khodjamirian:2011ub}, $f_{+}(0)=0.31\pm0.02$ \cite{Imsong:2014oqa} or the recent $f_{+}(0)=0.208\pm0.007\pm0.015\pm0.030$ \cite{Yao:2018tqn}.

\begin{figure}[h!]
\begin{center}
\includegraphics[scale=0.725]{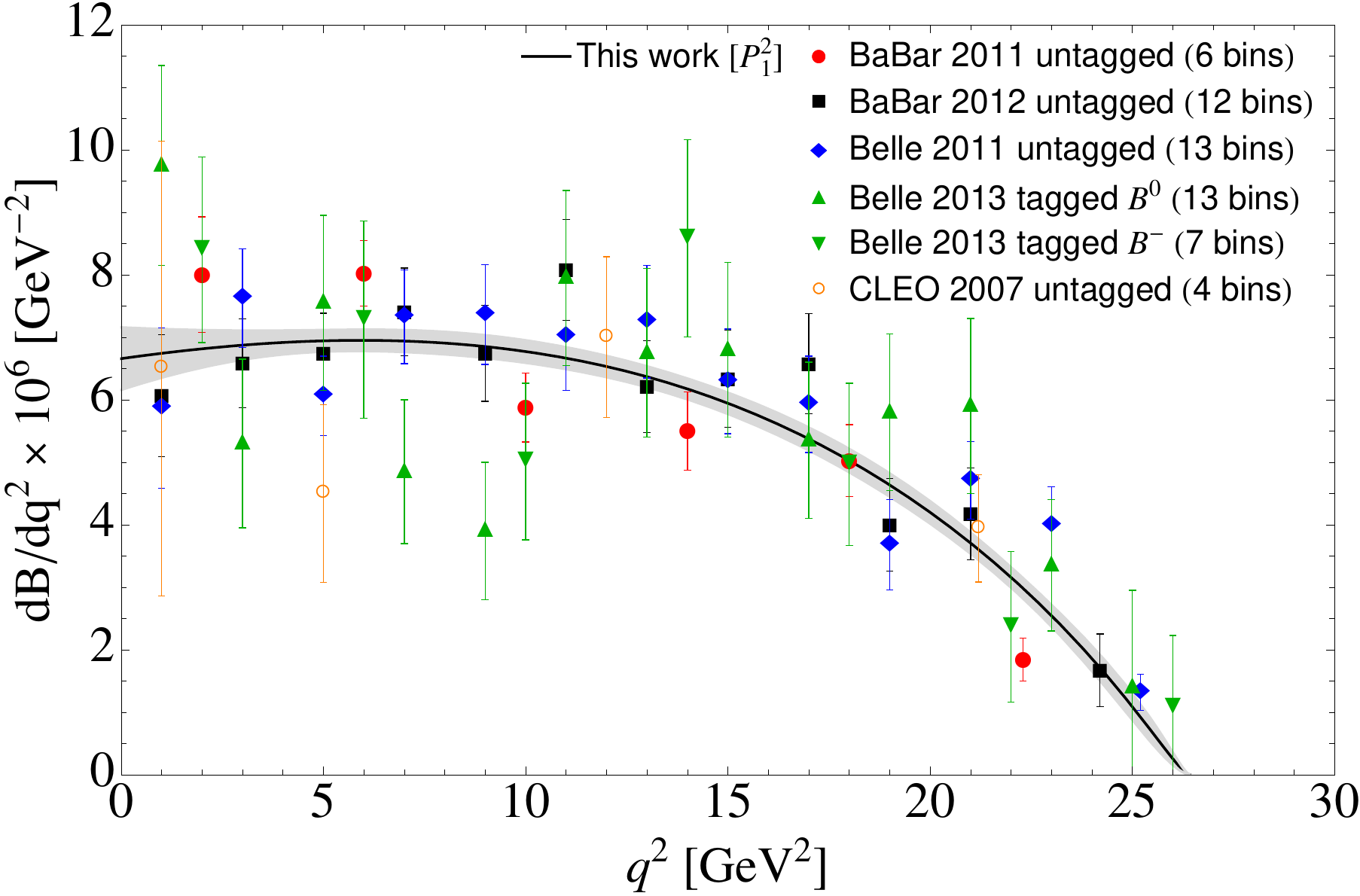}
\caption{\label{spectrum}Simultaneous fit to BaBar~\cite{delAmoSanchez:2010af,Lees:2012vv} and Belle~\cite{Ha:2010rf,Sibidanov:2013rkk} $B\to\pi\ell\nu_{\ell}$ experimental data as obtained from the $\chi^{2}_{\rm{data}}$ minimization of Eq.~(\ref{chi2data}) with a $P_{1}^{2}(q^{2})$ approximant (black solid line). 
CLEO data~\cite{Adam:2007pv} is not included from the fit and rather shown for comparison.}
\end{center}
\end{figure}

We start fitting with ordinary PAs of the type $P_{1}^{M}(q^{2})$ and $P_{2}^{M}(q^{2})$ where the poles are left free to be fitted and we reach $M=2$ and $M=0$, respectively.
Then, we proceed to fit with sequences of the type $T_{1}^{M}(q^{2})$ and $P_{1,1}^{M}(q^{2})$ by fixing the $B^{*}$ pole to $m_{B^{*}}=5.325$ GeV reaching, respectively, $M=2$ and $M=1$.
In Fig.~\ref{spectrum}, we provide a graphical account of the fit as obtained with $P_{1}^{2}(q^2)$ compared to data while our fit results are collected in the first row of Table~\ref{fitspectra}.
From the plot we observe that the uncertainty associated to the fit, given by the gray error band, is slightly larger in the low-$q^{2}$ energy region while from the Table we read that the values for $|V_{ub}|$ determined with approximants with two poles i.e. $P_{2}^{1}(q^2)$ and $P_{1,1}^{1}(q^2)$, give identical results than the single pole ones, $P_{1}^{2}(q^2)$ and $T_{1}^{2}(q^2)$.

\begin{table}[h!]
  \centering
  \begin{tabular}{|c|l|l|c|c|c|c|c|c|c|c|}
\hline
    \multirow{1}{*}{Fit} & 
    \multicolumn{2}{c|}{\multirow{1}{*}{Approximant}} & 
    \multicolumn{1}{c|}{$\chi^{2}_{\rm{dof}}$}& 
    \multicolumn{1}{c|}{Poles (GeV)}&
    \multicolumn{1}{c|}{$|V_{ub}f_{+}(0)|\times10^{4}$}&
    \multicolumn{1}{c|}{$|V_{ub}|\times10^{3}$}\\
    \cline{4-7}
    \hline
      \multirow{2}{*}{BaBar+Belle} & \multirow{2}{*}{Free $B^{*}$} &$P_{1}^{2}$&1.70&5.32&$9.27(37)$&$3.57(33)$\\ \cline{3-7}
  \multirow{2}{*}{\cite{delAmoSanchez:2010af,Lees:2012vv,Ha:2010rf,Sibidanov:2013rkk}} & & \multirow{1}{*}{$P_{2}^{1}$} & 1.71&$5.31,\,15.53^{\dagger}$&$9.24^{+0.29}_{-0.34}$&$3.56(35)$\\ \cline{2-7}
  & \multirow{2}{*}{Fixed $B^{*}$} & \multirow{1}{*}{$T_{1}^{2}$} &$1.70$&--&$9.28(31)$&$3.57(33)$\\ \cline{3-7}
    & & \multirow{1}{*}{$P_{1,1}^{2}$} &$1.57$&$9.83$&$9.26^{+0.34}_{-0.30}$&$3.56(35)$\\ \cline{2-7}
        \hline
    \hline
          \multirow{2}{*}{BaBar+Belle} & \multirow{2}{*}{Free $B^{*}$} &$P_{1}^{2}$&1.63&5.32&$9.25(37)$&$3.56(33)$\\ \cline{3-7}
          \multirow{2}{*}{+CLEO} & & \multirow{1}{*}{$P_{2}^{1}$} & 1.63&$5.32,\,13.86^{\dagger}$&$9.21^{+0.29}_{-0.33}$&$3.54(34)$\\ \cline{2-7}
 \multirow{2}{*}{\cite{Adam:2007pv,delAmoSanchez:2010af,Lees:2012vv,Ha:2010rf,Sibidanov:2013rkk}}  & \multirow{2}{*}{Fixed $B^{*}$} & \multirow{1}{*}{$T_{1}^{2}$} &$1.74$&--&$9.25(31)$&$3.56(32)$\\ \cline{3-7}
    & & \multirow{1}{*}{$P_{1,1}^{2}$} &$1.64$&$8.30$&$9.33^{+0.25}_{-0.45}$&$3.55(36)$\\ \cline{2-7}
        \hline
    \hline
\multirow{1}{*}{CLEO 07 \cite{Adam:2007pv}}   & \multirow{1}{*}{Fixed $B^{*}$} & \multirow{1}{*}{$P_{1,1}^{0}$} &$0.94$&$6.66$&$8.44(96)$&$3.25(27)$\\ \cline{1-7}
    \hline
    \hline
    \multirow{3}{*}{BaBar 11} & \multirow{2}{*}{Free $B^{*}$} &$P_{1}^{1}$&2.94&5.36&$10.49(50)$&$4.03(40)$\\ \cline{3-7}
 \multirow{3}{*}{\cite{delAmoSanchez:2010af}}  & & \multirow{1}{*}{$P_{2}^{0}$} & 2.81&$5.43,\,8.81$&$10.46(51)$&$4.02(39)$\\ \cline{2-7}
  & \multirow{2}{*}{Fixed $B^{*}$} & \multirow{1}{*}{$T_{1}^{2}$} &$2.93$&--&$10.74(69)$&$4.13(44)$\\ \cline{3-7}
    & & \multirow{1}{*}{$P_{1,1}^{1}$} &$2.94$&$10.37^{\dagger}$&$10.67^{+0.49}_{-0.69}$&$4.11(48)$\\ \cline{1-7}
      \multicolumn{5}{|l|}{\multirow{1}{*}{BaBar 11 reported value}} &$10.80(56)$&\\ \cline{2-7}
    \hline
    \hline
        \multirow{3}{*}{BaBar 12} & \multirow{2}{*}{Free $B^{*}$} &$P_{1}^{2}$&0.70&5.36&8.58(59)&3.30(36)\\ \cline{3-7}
  \multirow{3}{*}{\cite{Lees:2012vv}}  & & \multirow{1}{*}{$P_{2}^{2}$} &0.81&$5.35, \, 15.12^{\dagger}$&$8.59^{+0.62}_{-0.47}$&3.30(41)\\ \cline{2-7}
  & \multirow{2}{*}{Fixed $B^{*}$} & \multirow{1}{*}{$T_{1}^{3}$} &$0.70$&--&$8.62^{+0.73}_{-0.76}$&3.32(49)\\ \cline{3-7}
    & & \multirow{1}{*}{$P_{1,1}^{2}$} &$0.70$&$13.87^{\dagger}$&$8.54^{+0.56}_{-0.54}$&3.29(41)\\ \cline{1-7}
    \multicolumn{5}{|l|}{\multirow{1}{*}{BaBar 12 reported value}} &$8.7(3)$&\\ \cline{2-7}
    \hline
    \hline
    \multirow{3}{*}{Belle 11} & \multirow{2}{*}{Free $B^{*}$} &$P_{1}^{2}$&1.32&5.25&$8.91^{+0.60}_{-0.61}$&3.43(27)\\ \cline{3-7}
    \multirow{3}{*}{\cite{Ha:2010rf}} & & \multirow{1}{*}{$P_{2}^{3}$} &1.68&$5.25^{\dagger\dagger}$&$9.05(70)$&3.48(48)\\ \cline{2-7}
  & \multirow{2}{*}{Fixed $B^{*}$} & \multirow{1}{*}{$T_{1}^{2}$} &$1.34$&--&9.19(57)&3.53(37)\\ \cline{3-7}
    & & \multirow{1}{*}{$P_{1,1}^{2}$} &$1.31$&$5.35^{\dagger\dagger}$&$9.02^{+0.46}_{-0.44}$&3.47(38)\\ \cline{1-7}
    \multicolumn{5}{|l|}{\multirow{1}{*}{Belle 11 reported value}} &$9.23(18)_{\rm{stat}}(21)_{\rm{syst}}$&\\ \cline{2-7}
    \hline    
    \hline
    \multirow{3}{*}{Belle 13} & \multirow{2}{*}{Free $B^{*}$} &$P_{1}^{1}$&1.43&5.19&$9.13^{+0.58}_{-0.62}$&3.51(44)\\ \cline{3-7}
  \multirow{3}{*}{\cite{Sibidanov:2013rkk}} & & \multirow{1}{*}{$P_{2}^{0}$}&1.34&$5.26,\,7.38$&9.19(53)&3.54(36)\\ \cline{2-7}
  & \multirow{2}{*}{Fixed $B^{*}$} & \multirow{1}{*}{$T_{1}^{1}$} &$1.51$&--&$8.68^{+0.59}_{-0.61}$&3.34(43)\\ \cline{3-7}
    & & \multirow{1}{*}{$P_{1,1}^{1}$} &$1.31$&$5.96$&$9.44^{+0.53}_{-0.59}$&3.63(43)\\ \cline{2-7}
    \hline        
  \end{tabular}
  \caption{\label{fitspectra}The product $|V_{ub}f_{+}(0)|$ as obtained from fits to $B\to\pi\ell\nu_{\ell}$ data depending if the $B^{*}$ pole is let as a free parameter to fit or fixed at $m_{B^{*}}=5.325$ GeV. The corresponding $|V_{ub}|$ value extracted using $f_{+}(0)=0.261^{+0.020}_{-0.023}$~\cite{Bharucha:2012wy}, the pole(s) of the approximants and the $\chi^2_{\rm{dof}}$ are also shown. 
Poles placed far away from the origin and \textit{Froissart doublet} are denoted by $^{\dagger}$ and $^{\dagger\dagger}$, respectively.  
Errors are only statistical and symmetrized in the last column.}
\end{table}

Then, we add CLEO 2007 experimental data into the $\chi^{2}$ minimization of Eq.~(\ref{chi2data}) and report the corresponding fit results in the second row of the Table~\ref{fitspectra}, which upon comparison with the results shown in the first row we conclude that the effect of including this data into the fit is tiny. 

In order to further improve on what can be learned from experimental data, we have also fitted experimental data of each Collaboration separately, an exercise that will be very illustrative in order to determine $|V_{ub}|$.

\begin{figure}[h!]
\begin{center}
\includegraphics[scale=0.3]{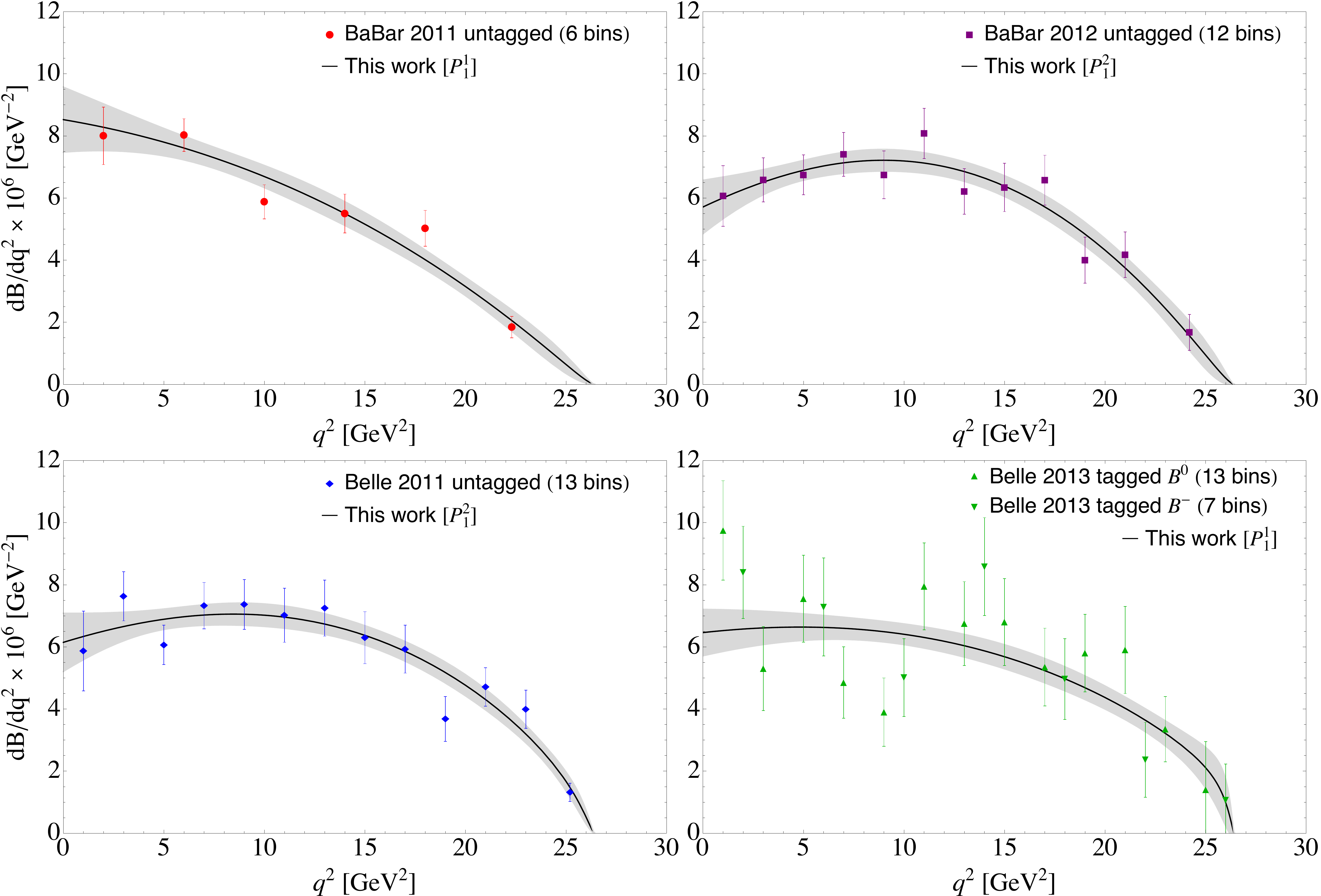}
\caption{\label{distcoll}Individual fits to BaBar11~\cite{delAmoSanchez:2010af} (left-top panel), BaBar12~\cite{Lees:2012vv} (right-top panel), Belle11~\cite{Ha:2010rf} (left-down panel) and Belle13~\cite{Sibidanov:2013rkk} (right-down panel) experimental data sets on the $B\to\pi\ell\nu_{\ell}$ differential branching ratio distribution.}
\end{center}
\end{figure}

The individual fits are displayed in Fig.~\ref{distcoll} and the corresponding results shown, respectively, in the third (CLEO07), fourth (BaBar11), fifth (BaBar12), sixth (Belle11) and seventh (Belle13) rows of Table~\ref{fitspectra}.
From our set of fits collected in this Table, the diagonal and near diagonal $P_{2}^{2}(q^2)$, $P_{1,(1)}^{1}(q^2)$ and $P_{1,(1)}^{2}(q^2)$ approximants deserve special attention.
For these approximants we find some extraneous poles placed either far away from the origin (marked with $^{\dagger}$ in the Table) or pair up with a close-by zero in the numerator becoming what is called a \textit{defect} or \textit{Froissart doublet} (marked with $^{\dagger\dagger}$ in the Table), in accordance with the Nutall-Pommerenke's convergence theorem~\cite{Masjuan:2007ay,Baker}.
We would like to point out that the zeros of the numerator when individual fits to the BaBar 2012 data are performed tend to lie within the radius of convergence in the region of negative $q^{2}$, a region we expected out of zeros.
This feature may explain why the corresponding distribution is more rounded and with a sizable negative fall off at the origin in comparison with the other three individual fits.
We also note that a \textit{Froissart doublet} doublet appears at $q^{2}=-1$ GeV for the $P_{2}^{3}$ approximant reached for individual fits to the Belle data of 2011.
Let us remind that a PA to a Stieltjes function is also a Stieltjes function as well~\cite{Baker}. 
As such, it must have an imaginary part positive defined. 
This feature does not correspond with what we obtain in our fits and disagrees with what we expect from a Stieltjes function. 
All zeros and poles of our approximants must lie along the unitary branch cut in order to fulfill the unitary requirements that the FF imposes. 
If a particular PA does not show this feature means the set of data fitted is not fulfilling the unitary requirements that must have. 
Thus, both defects and the appearance of poles and zeros outside the unitary branch cut are indications of a violation of unitary to a certain degree. 
We shall come back to this point later (see section \ref{unitaryconstraints}).

We would like to notice that individual fits to CLEO data lead unrealistic results but for $P_{1,1}^{0}(q^2)$.
Also notice that the fits to BaBar11 experimental data lead the worst $\chi^{2}/\rm{dof}$, in agreement with Ref.~\cite{Lattice:2015tia}, and the largest values for $|V_{ub}|$, in line with Ref.~\cite{Bharucha:2012wy} but in contradiction with Refs.~\cite{Lattice:2015tia,Flynn:2015mha}.
These two features are somehow reflected in the left-top panel of Fig.~\ref{distcoll} both in the error band and in the value of the branching ratio distribution at $q^{2}=0$ which are, respectively, wider and larger than in the other three panels of the figure.
On the contrary, the fits to the BaBar12 data gives the best $\chi^{2}/\rm{dof}$ and tends to give smaller $|V_{ub}|$ values.

From each of the individual fits shown in Table~\ref{fitspectra} we can order the experimental Collaborations according to their bottom-up $|V_{ub}|$ values as: BaBar12, Belle11, Belle13 and BaBar11. This ordering is in line with the corresponding $|V_{ub}|$ values reported by the experimental groups from fits to their own experimental data. 

\begin{figure}[h!]
\begin{center}
\includegraphics[scale=0.4]{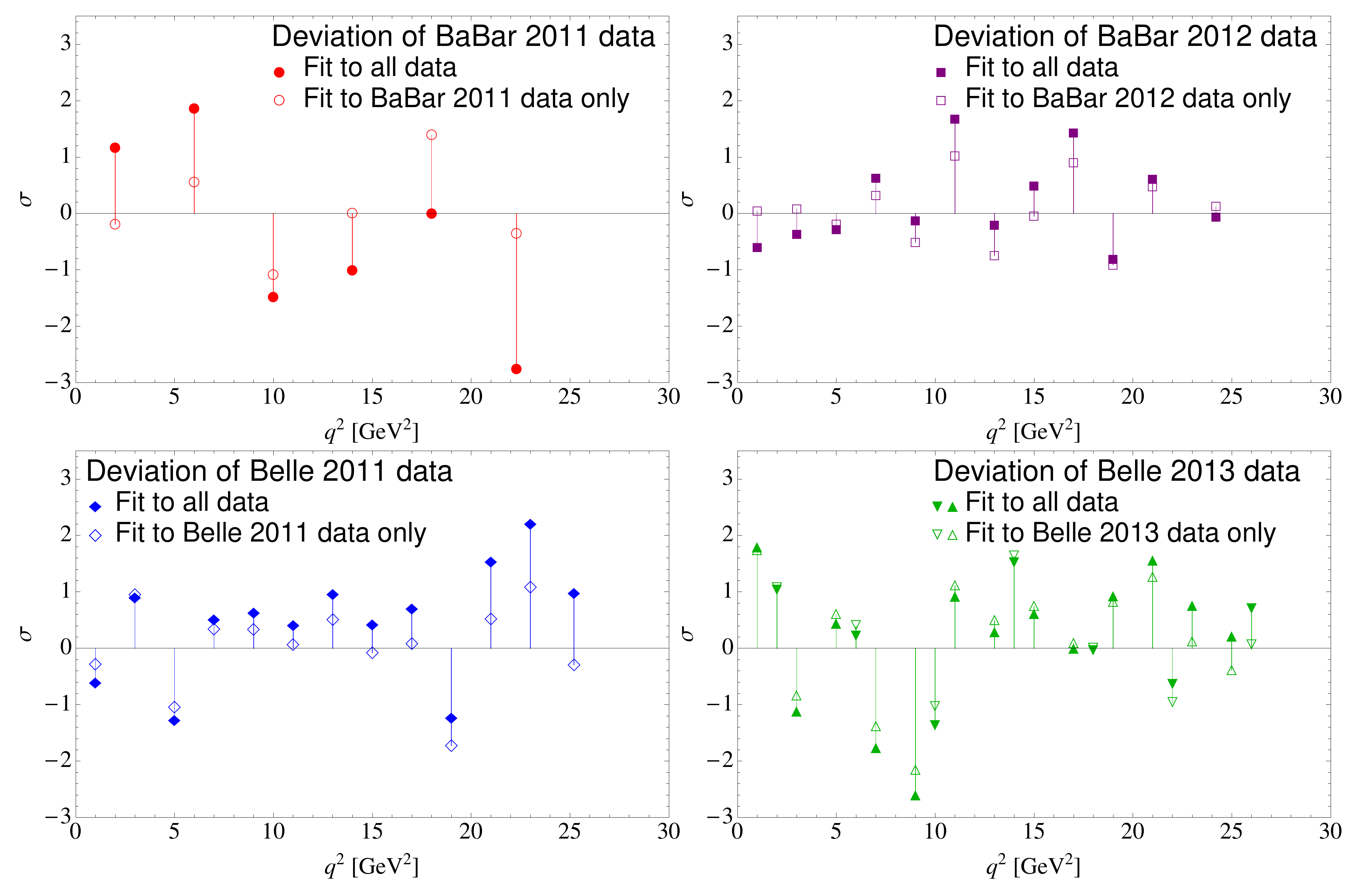}
\caption{\label{sigmapoints}Deviation, in $\sigma$, of each experimental datum with respect to our combined and individual fits. Solid and empty geometric markers account, respectively, for the fits as given in Figs.~\ref{spectrum} and \ref{distcoll}.}
\end{center}
\end{figure}

The neat effect of fitting all experimental data sets together with respect to fitting data of each Collaboration separately can be seen in Fig.~\ref{sigmapoints}, where we represent the number of $\sigma$ deviations of each experimental datum with respect the corresponding fits. 
In this figure, markers given by solid geometric figures accounts for the fit as given in Fig.~\ref{spectrum} while empty geometric figures stand for the fits as shown in Fig.~\ref{distcoll}. 
This allows us to order the four experimental data sets according to their increasing degree of soundness with respect to the common fit i.e., BaBar11, Belle13, Belle11 and BaBar12. Clearly, BaBar11 data points suffer the largest deviation when including the other sets of data into the fit (see left-top panel in Fig.~\ref{sigmapoints}) while, on the contrary, BaBar12 and Belle11 data points seem to drive the $\chi^{2}$ minimization dominating the fit (see right-top and left-down panels, respectively, in Fig.~\ref{sigmapoints}).
Regarding Belle13 experimental data points, they show some oscillatory scatter lying in between BaBar11 and BaBar12/Belle11 cases.

\subsection{Incorporating form factor lattice calculations}\label{section4}

In the previous section we have not accessed the description of the form factor but rather its normalized version to unity at $q^{2}=0$.
In order to achieve a parameterization of the form factor we include the form factor shape predictions at large $q^{2}$ obtained on the lattice as new data sets to be fitted.
In particular, we consider Table V of the HPQCD Collaboration results~\cite{Dalgic:2006dt} (consider the erratum version of it), Table II of the 2008 MILC Collaboration predictions~\cite{Bailey:2008wp}, Table VI of the RBC and UKQCD Collaboration calculation~\cite{Flynn:2015mha} and, finally, the updated analysis of the FNAL/MILC Collaboration of 2015~\cite{Lattice:2015tia}\footnote{\label{latticedata}While the FF predictions as obtained by HPQCD2007, MILC2008 and RBC/UKQCD in Refs.~\cite{Dalgic:2006dt,Bailey:2008wp,Flynn:2015mha}, respectively, are publicly available and the corresponding results given in the papers, the updated 2015 results of MILC are not. 
However, we have generated the FF from the fit as given in Table XIV of Ref.~\cite{Lattice:2015tia}.
For the sake of comparison with their former 2008 predictions, we have generated 12 data points placed at the same $q^{2}$-bins.
For the extracted data points, the interested reader may contact the corresponding authors. 
We would like to thank Elvira G\'{a}miz for correspondence along these lines.}.   
The main advantage of performing a simultaneous fit to all measured $q^{2}$ spectra experimental data supplemented by lattice QCD results on the FF shape is that not only $|V_{ub}|$ but also $f_{+}(0)$ can be determined directly from the fit since lattice data drive the height of the curve of the decay spectra.
The $\chi^{2}$ function to be minimized reads
\begin{eqnarray}
\chi^{2}&=&\chi^{2}_{\rm{data}}+\sum_{i=1}^{5}\left(\frac{f_{+}^{{\rm{HPQCD}07}}(q^{2})_{i}-P_{N}^{M}(q^{2})_{i}}{\sigma_{i}^{{\rm{HPQCD}07}}}\right)^{2}+\sum_{i=1}^{12}\left(\frac{f_{+}^{{\rm{MILC08}}}(q^{2})_{i}-P_{N}^{M}(q^{2})_{i}}{\sigma_{i}^{{\rm{MILC08}}}}\right)^{2}\nonumber\\
&+&\sum_{i=1}^{3}\left(\frac{f_{+}^{{\rm{UKQCD15}}}(q^{2})_{i}-P_{N}^{M}(q^{2})_{i}}{\sigma_{i}^{{\rm{UKQCD15}}}}\right)^{2}+\sum_{i=1}^{12}\left(\frac{f_{+}^{{\rm{MILC15}}}(q^{2})_{i}-P_{N}^{M}(q^{2})_{i}}{\sigma_{i}^{{\rm{MILC15}}}}\right)^{2}\,,
\label{chi2}
\end{eqnarray}
where $\chi^{2}_{\rm{data}}$ corresponds to Eq.~(\ref{chi2data}) and $\sigma_{i}$ is the uncertainty corresponding to the $i$-th bin.

The results derived from the minimization of Eq.~(\ref{chi2}) are collected in Table~\ref{fitspectralattice}.
In contrast to Table~\ref{fitspectra}, Table~\ref{fitspectralattice} collects the results as obtained with each element of the corresponding sequences going up to $P_{1}^{2},P_{2}^{2},T_{1}^{2}$ and $P_{1,1}^{2}$, respectively. 
The final results, given in the last column, include both statistical, from the fit, and systematic uncertainties from the truncated PA sequence as the difference of central values of the element we have stopped the sequence and the preceding one.
Notice that the systematic uncertainty increases when the $B^{*}$ pole is fixed.

The impact of including lattice data into the fit is evident and allow us to determine $|V_{ub}|$ with improved precision reducing the associated statistical uncertainty by $\sim80\%$ with respect to the case when only the decay spectra is fitted (cf.~Table~\ref{fitspectra}).
In addition, the sequence $P_{2}^{L}$ has been enlarged by one element. 
Again, we find some extraneous poles for the diagonal $P_{2}^{2}$ and $P_{1,1}^{2}$ elements.
In the former, we find a complex-conjugate (c.c.) pole with a small imaginary part (see the dedicated discussion in section \ref{unitaryconstraints}) while in the latter we find that one pole tends to pair up with a close-by zero in the numerator. 

\begin{figure}[h!]
\begin{center}
\includegraphics[scale=0.725]{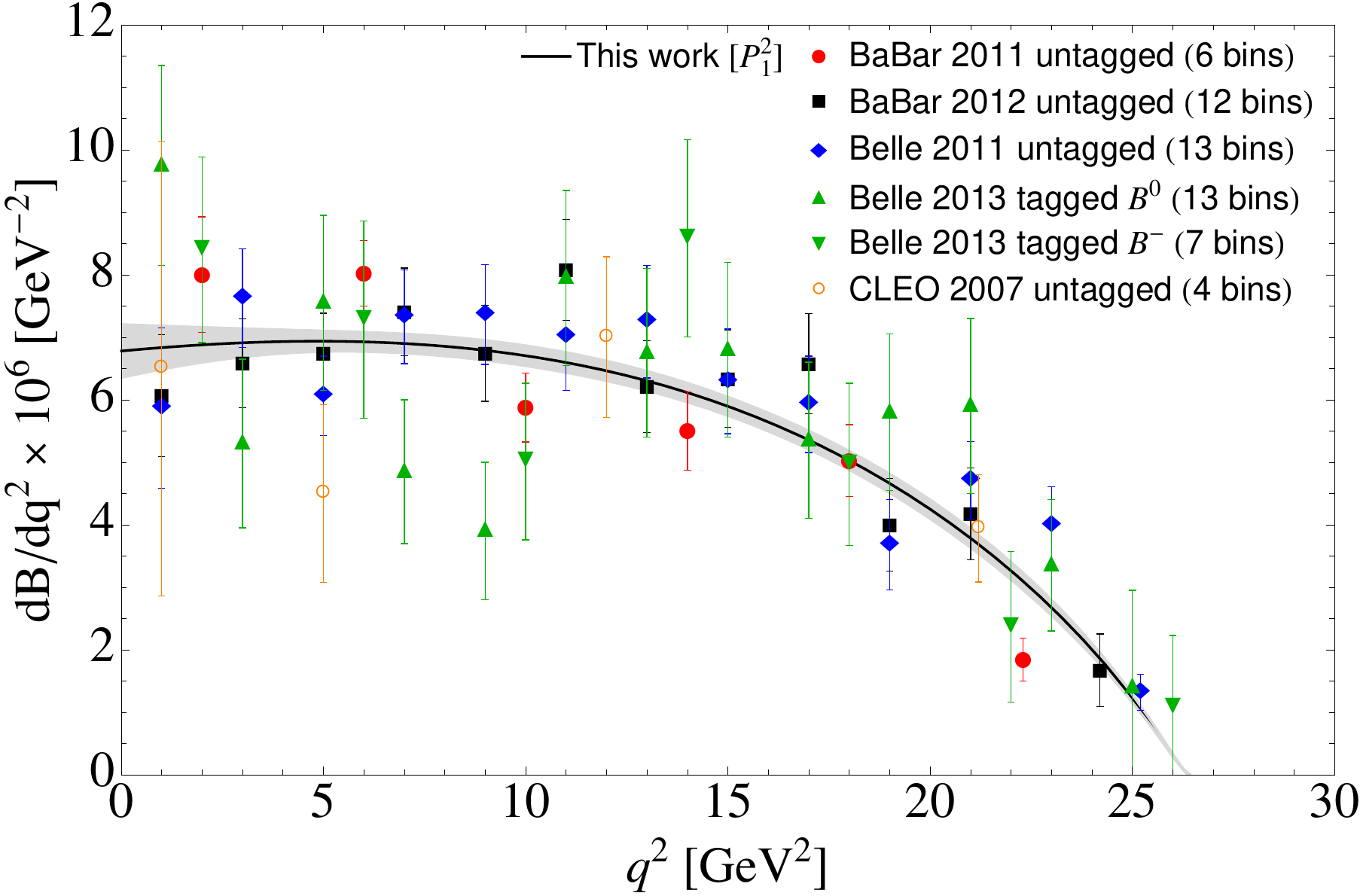}
\caption{\label{spectra_joint_fit}Differential branching ratio distribution for $B\to\pi\ell\nu_{\ell}$ decays as obtained from a combined fit to experimental data and lattice predictions on the form factor shape with a $P_{1}^{2}(q^{2})$ (black solid curve). CLEO data~\cite{Adam:2007pv} is excluded from the fit but rather shown for comparison.}
\end{center}
\end{figure}
\begin{figure}[h!]
\begin{center}
\includegraphics[scale=0.725]{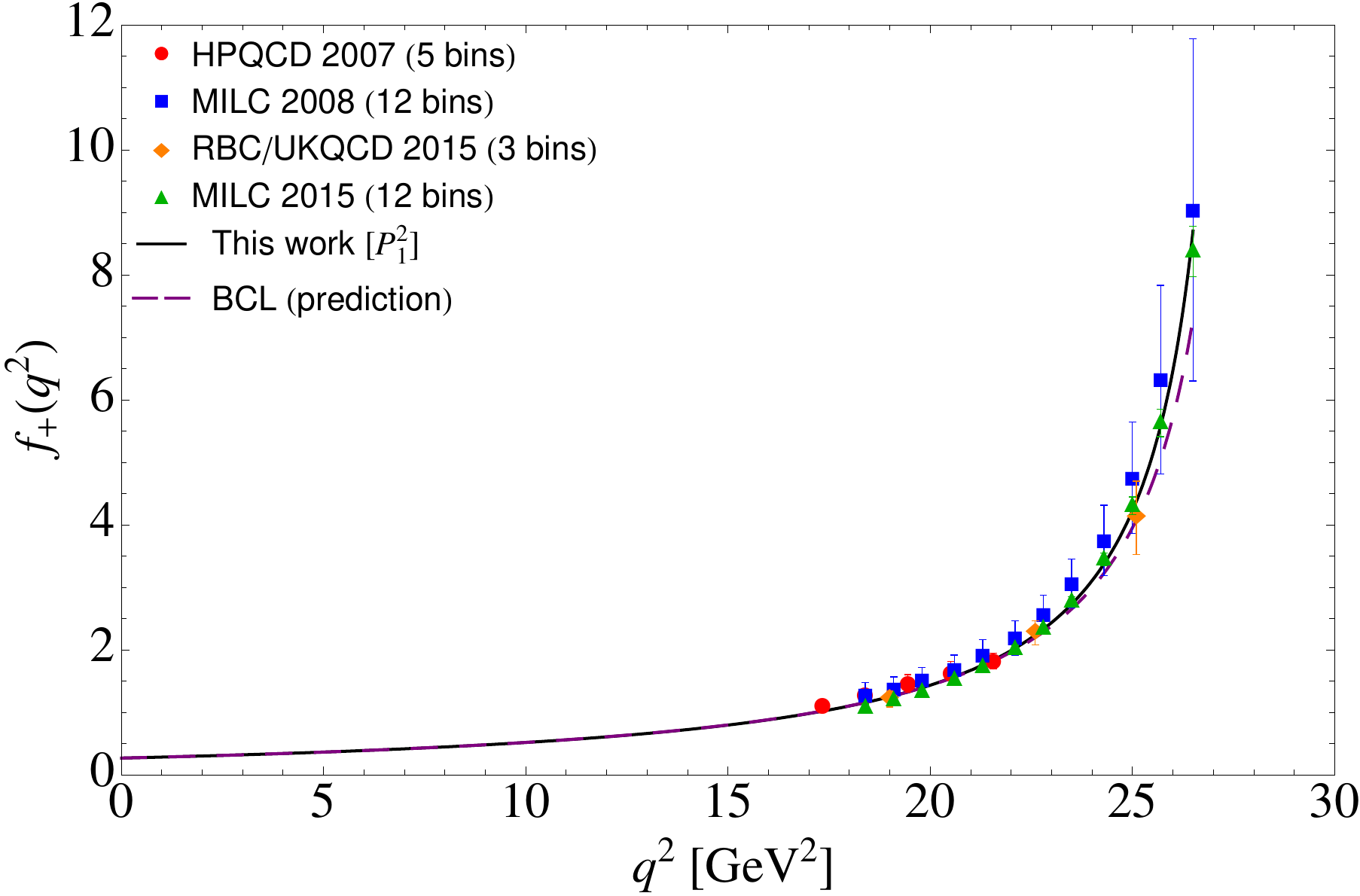}
\caption{\label{FF_joint_fit}$B\to\pi$ form factor as obtained from a combined fit to experimental data and lattice predictions on the form factor shape with the approximant $P_{1}^{2}(q^{2})$ (black solid curve). 
Our prediction for the BCL parameterization is also shown (purple dashed curve).}
\end{center}
\end{figure}

As a matter of example, we gather the coefficients of the Pad\'{e} approximant $P_{1}^{2}(q^{2})$ in Table~\ref{fittedcoefficients}. 
In this Table we also provide the series coefficients $b_{+}^{(n)}$ corresponding to the BCL parameterization (cf.\,Eq.\,(\ref{zparam2})) that are obtained by matching the Taylor series expansion of $P_{1}^{2}(q^{2})$ with the power series expansion of the BCL parameterization at $\mathcal{O}(q^{4})$. 
The coefficients thus obtained are not directly fitted to data but rather reconstructed from our rational function. 
These lie in the ballpark of the most recent RBC/UKQCD and FNAL/MILC lattice determinations \cite{Flynn:2015mha,Lattice:2015tia} shown, respectively, in the fifth and sixth columns of the Table and are seen in nice agreement with the HFLAV fit values \cite{Amhis:2016xyh} given in the last column.
A graphical account of the corresponding $P_{1}^{2}(q^{2})$ combined fit result is depicted in Figs.~\ref{spectra_joint_fit} and \ref{FF_joint_fit} as compared to the decay spectra and FF lattice data, respectively.
In the latter, our prediction for the BCL parameterization is also shown (purple dashed curve), accommodating pretty well all lattice data but the last datum and seen in nice agreement with the $P_{1}^{2}(q^{2})$ (black solid curve) element it is reconstructed from.
A closer look to the FF shape displayed in Fig.~\ref{FF_joint_fit} reveals that the lattice simulations derived by the FNAL/MILC Collaboration in 2015 seem to dominate the large $q^{2}$ region (cf.~Table~\ref{fitresultsvslattice}).

Our preferred values for $|V_{ub}|$ and $f_{+}(0)$ after the simultaneous fit results shown in Table~\ref{fitspectralattice} is
\bea
|V_{ub}|=3.53(8)_{\rm{stat}}(3)_{\rm{syst}}\times10^{-3}\,,\quad f_{+}(0)=0.265(10)_{\rm{stat}}(2)_{\rm{syst}}\,,
\label{centralresults1}
\eea
corresponding to $P_{1}^{2}(q^{2})$ when the pole is let as a free parameter to fit. 
This choice is based on the fact that the second pole of the sequence of the type $P_{2}^{M}$ is rare indicating that the single pole behavior for the form factor seems favored. 

To compare on the same footing regarding the number of free parameters we choose 
\bea
|V_{ub}|=3.53(8)_{\rm{stat}}(5)_{\rm{syst}}\times10^{-3}\,,\quad f_{+}(0)=0.264(10)_{\rm{stat}}(5)_{\rm{syst}}\,,
\label{centralresults2}
\eea
corresponding to the partial Pad\'{e} $P_{1,1}^{2}$ when the $B^{*}$ pole is fixed.
Notice that the corresponding systematic uncertainties are large enough to cover the difference with $P_{2}^{2}$ and $T_{1}^{2}$, respectively.

\begin{table}[h!]
\centering
\begin{tabular}{|l|llrrl|c|ccccccc}
\cline{1-6}
\multicolumn{6}{c}{Fits to BaBar and Belle data \cite{delAmoSanchez:2010af,Lees:2012vv,Ha:2010rf,Sibidanov:2013rkk}+Lattice FF predictions \cite{Lattice:2015tia,Dalgic:2006dt,Bailey:2008wp,Flynn:2015mha}}\\\cline{1-6}
Parameter&\multicolumn{4}{c}{Elements of the PA sequence}&\\\cline{1-6}
\hline
\hline
Free $B^{*}$ pole&\multicolumn{1}{l}{$P_{1}^{0}$}&\multicolumn{1}{l}{$P_{1}^{1}$}&\multicolumn{1}{l}{$P_{1}^{2}$}&\multicolumn{1}{l}{}&{\bf{Final result}}\\\cline{1-6}
$|V_{ub}|\times10^{3}$&\multicolumn{1}{l}{$2.76(5)_{\rm{stat}}$}&\multicolumn{1}{c}{$3.50(7)_{\rm{stat}}$}&\multicolumn{1}{l}{$3.53(8)_{\rm{stat}}$}&\multicolumn{1}{l}{}&$3.53(8)_{\rm{stat}}(3)_{\rm{syst}}$\\\cline{1-6}
$f_{+}(0)$&\multicolumn{1}{l}{$0.398(6)_{\rm{stat}}$}&\multicolumn{1}{c}{$0.263(9)_{\rm{stat}}$}&\multicolumn{1}{l}{$0.265(10)_{\rm{stat}}$}&\multicolumn{1}{l}{}&$0.265(10)_{\rm{stat}}(2)_{\rm{syst}}$\\\cline{1-6}
Pole (GeV)&\multicolumn{1}{l}{$5.26$}&\multicolumn{1}{l}{$5.29$}&\multicolumn{1}{l}{$5.29$}&\multicolumn{1}{l}{}&\\\cline{1-6}
$\chi^{2}_{\rm dof}$&\multicolumn{1}{l}{$3.25$}&\multicolumn{1}{l}{$1.26$}&\multicolumn{1}{l}{$1.27$}&\multicolumn{1}{l}{}&\\\cline{1-6}
\hline
\hline
Free $B^{*}$ pole&\multicolumn{1}{l}{$P_{2}^{0}$}&\multicolumn{1}{l}{$P_{2}^{1}$}&\multicolumn{1}{l}{$P_{2}^{2}$}&\multicolumn{1}{l}{}&{\bf{Final result}}\\\cline{1-6}
$|V_{ub}|\times10^{3}$&\multicolumn{1}{l}{$3.56(8)_{\rm{stat}}$}&\multicolumn{1}{c}{$3.53(9)_{\rm{stat}}$}&\multicolumn{1}{l}{$3.52(8)_{\rm{stat}}$}&\multicolumn{1}{l}{}&$3.52(8)_{\rm{stat}}(1)_{\rm{syst}}$\\\cline{1-6}
$f_{+}(0)$&\multicolumn{1}{l}{$0.270(8)_{\rm{stat}}$}&\multicolumn{1}{c}{$0.265(9)_{\rm{stat}}$}&\multicolumn{1}{l}{$0.263(9)_{\rm{stat}}$}&\multicolumn{1}{l}{}&$0.263(9)_{\rm{stat}}(2)_{\rm{syst}}$\\\cline{1-6}
Poles (GeV)&\multicolumn{1}{l}{$5.30\&7.63$}&\multicolumn{1}{l}{$5.29\&11.24^{\dagger}$}&\multicolumn{1}{l}{c.c.}&\multicolumn{1}{l}{}&\\\cline{1-6}
$\chi^{2}_{\rm dof}$&\multicolumn{1}{l}{$1.27$}&\multicolumn{1}{l}{$1.27$}&\multicolumn{1}{l}{$1.31$}&\multicolumn{1}{l}{}&\\\cline{1-6}
\hline
\hline
Fixed $B^{*}$ pole&\multicolumn{1}{l}{$T_{1}^{0}$}&\multicolumn{1}{l}{$T_{1}^{1}$}&\multicolumn{1}{l}{$T_{1}^{2}$}&\multicolumn{1}{l}{}&{\bf{Final result}}\\\cline{1-6}
$|V_{ub}|\times10^{3}$&\multicolumn{1}{l}{$2.46(4)_{\rm{stat}}$}&\multicolumn{1}{c}{$3.42(7)_{\rm{stat}}$}&\multicolumn{1}{l}{$3.56(8)_{\rm{stat}}$}&\multicolumn{1}{l}{}&$3.56(8)_{\rm{stat}}(14)_{\rm{syst}}$\\\cline{1-6}
$f_{+}(0)$&\multicolumn{1}{l}{$0.450(4)_{\rm{stat}}$}&\multicolumn{1}{c}{$0.261(9)_{\rm{stat}}$}&\multicolumn{1}{l}{$0.272(9)_{\rm{stat}}$}&\multicolumn{1}{l}{}&$0.272(9)_{\rm{stat}}(11)_{\rm{syst}}$\\\cline{1-6}
$\chi^{2}_{\rm dof}$&\multicolumn{1}{l}{$4.63$}&\multicolumn{1}{l}{$1.52$}&\multicolumn{1}{l}{$1.36$}&\multicolumn{1}{l}{}&\\\cline{1-6}
\hline
\hline
Fixed $B^{*}$ pole&\multicolumn{1}{l}{$P_{1,1}^{0}$}&\multicolumn{1}{l}{$P_{1,1}^{1}$}&\multicolumn{1}{l}{$P_{1,1}^{2}$}&&{\bf{Final result}}\\\cline{1-6}
$|V_{ub}|\times10^{3}$&\multicolumn{1}{l}{$3.57(8)_{\rm{stat}}$}&\multicolumn{1}{c}{$3.58(8)_{\rm{stat}}$}&\multicolumn{1}{l}{$3.53(8)_{\rm{stat}}$}&&$3.53(8)_{\rm{stat}}(5)_{\rm{syst}}$\\\cline{1-6}
$f_{+}(0)$&\multicolumn{1}{l}{$0.266(8)_{\rm{stat}}$}&\multicolumn{1}{c}{$0.269(9)_{\rm{stat}}$}&\multicolumn{1}{l}{$0.264(10)_{\rm{stat}}$}&&$0.264(10)_{\rm{stat}}(5)_{\rm{syst}}$\\\cline{1-6}
Pole (GeV)&\multicolumn{1}{l}{$7.34$}&\multicolumn{1}{l}{$6.88$}&\multicolumn{1}{l}{$5.59^{\dagger\dagger}$}&&\\\cline{1-6}
$\chi^{2}_{\rm dof}$&\multicolumn{1}{l}{$1.29$}&\multicolumn{1}{l}{$1.30$}&\multicolumn{1}{l}{$1.30$}&&\\\cline{1-6}
\end{tabular}
\caption{$|V_{ub}|$ and $f_{+}(0)$ values as obtained from a simultaneous fit to $B\to\pi\ell\nu_{\ell}$ decay data and lattice QCD form factor simulations. The pole(s) of the approximants and the $\chi^{2}_{\rm dof}$ are also shown. 
Poles placed far away from the origin and \textit{Froissart doublet} are denoted by $^{\dagger}$ and $^{\dagger\dagger}$, respectively, while c.c. stands for a complex-conjugate pole with a small imaginary part.
The results in the last column include a systematic error coming from the difference of central values of the last two elements of the corresponding PA sequences. The errors are symmetrized.}
\label{fitspectralattice}
\end{table}

\begin{table}[h!]
\centering
\begin{tabular}{|cl||cllcl|}
\hline
Pad\'{e}&$P_{1}^{2}(q^2)$& BCL& our prediction& UKQCD\,\cite{Flynn:2015mha}&MILC\,\cite{Lattice:2015tia}&HFLAV\,\cite{Amhis:2016xyh}\\
\hline
$a_0$ & $0.265(10)$& $b_{+}^{(0)}$& $0.424(78)$& $0.412(39)$&$0.419(13)$&0.418(12)\\ 
$a_1$ & $0.006(2)$& $b_{+}^{(1)}$& $-0.343(199)$& $-0.511(184)$&$-0.495(55)$&-0.399(33)\\
$a_2$ & $0.00006(8)$& $b_{+}^{(2)}$& $-0.632(368)$& $-0.524(612)$&$-0.43(14)$&-0.578(130)\\
$b_1$ & $-0.0357(1)$&&&&&\\
\cline{1-7}                                           
\end{tabular}
\caption{Coefficients of the Pad\'e approximant $P_{1}^{2}(q^{2})$, with the pole let as a free parameter, and of the reconstructed BCL parameterization, where the pole is fixed to the $B^{*}$. The latter are compared with the fitted coefficients determined by the RBC/UKQCD and FNAL/MILC lattice groups \cite{Flynn:2015mha,Lattice:2015tia} and with the HFLAV results \cite{Amhis:2016xyh}.}
\label{fittedcoefficients}
\end{table}

The impact of including the value of the FF at $q^{2}=0$, $f_{+}(0)=0.261^{+0.020}_{-0.023}$~\cite{Bharucha:2012wy}, as an external restriction in the $\chi^{2}$, Eq.~(\ref{chi2}), is probed through the fits displayed in Table~\ref{fitspectralatticeconstrainedf0} for our best fit sequences discussed previously when $f_{+}(0)$ was not included in the minimization (cf.~Table~\ref{fitspectralattice}), and repeated in this Table~\ref{fitspectralatticeconstrainedf0}, second column, for ease of comparison. The corresponding fits are almost identical, which guarantees the independency of our results with respect of the model calculation of $f_{+}(0)$.


\begin{table}[h!]
\centering
\begin{tabular}{lcccccccccccccc}
\hline
\multicolumn{1}{l}{\bf{Free} $B^{*}$ pole}& 
\multicolumn{2}{c}{Table~\ref{fitspectralattice}}& 
\multicolumn{2}{c}{\bf{Constraining} $f_{+}(0)$}\\[1ex]
&\textbf{$P_{1}^{2}$}&\textbf{$P_{2}^{2}$}& \textbf{$P_{1}^{2}$}&\textbf{$P_{2}^{2}$}\\
\hline
$|V_{ub}|\times10^{3}$&$3.53(8)(3)$&$3.52(9)(1)$&3.52(9)(5)&3.51(9)(1)\\[0.5ex]
$f_{+}(0)$&$0.265(10)(2)$&$0.265(9)(2)$&0.264(8)(4)&0.264(8)(2)\\[0.5ex]
Pole(s) (GeV)&5.29&$5.30, \,16.39$&$5.30$&$5.30, \,18.07$\\[0.5ex]
$\chi^{2}_{\rm dof}$&1.27&1.31&1.30&1.31\\[0.5ex]
\hline
\multicolumn{1}{l}{\bf{Fixed} $B^{*}$ pole}& 
\multicolumn{2}{c}{Table~\ref{fitspectralattice}}& 
\multicolumn{2}{c}{\bf{Constraining} $f_{+}(0)$}\\[1ex]
&\textbf{$T_{1}^{2}$}&\textbf{$P_{1,1}^{2}$}& \textbf{$T_{1}^{2}$}&\textbf{$P_{1,1}^{3}$}\\
\hline
$|V_{ub}|\times10^{3}$&$3.56(8)(14)$&$3.53(8)(5)$&3.54(8)(14)&3.50(9)(2)\\[0.5ex]
$f_{+}(0)$&$0.272(9)(11)$&$0.264(10)(5)$&0.268(8)(9)&0.258(9)(4)\\[0.5ex]
Pole (GeV)&---&$5.59$&---&$5.88$\\[0.5ex]
$\chi^{2}_{\rm dof}$&1.36&1.30&1.36&1.26\\[0.5ex]
\hline
\end{tabular}
\caption{Comparison between the $|V_{ub}|$ and $f_{+}(0)$ values as obtained from a simultaneous fit to $B\to\pi\ell\nu_{\ell}$ experimental data and lattice predictions on the FF shape with (second multicolumn) and without (first multicolumn) the LCSR prediction $f_{+}(0)=0.261^{+0.020}_{-0.023}$~\cite{Bharucha:2012wy} as external restriction into the fit. 
In the upper part of the Table, the $B^{*}$ pole is left free to fit while in the lower ones the pole is fixed. The pole(s) of the approximants and the $\chi^{2}_{\rm dof}$ are also shown. The first and second parenthesis refer, respectively, to the statistical and systematic uncertainties. The errors are symmetrized.}
\label{fitspectralatticeconstrainedf0}
\end{table}

We have also performed fits including CLEO 2007 data~\cite{Adam:2007pv} in the $\chi^{2}$, and found that their impact in the global fit is marginal and we hence refrain to show them, and explored the effect of fitting all experimental data together with certain groups of lattice FF simulations.
We have considered three groups, HPQCD+RBC/UKQCD, HPQCD+RBC/UKQCD+MILC08 and HPQCD+RBC/UKQCD+MILC2015, and collected the results, respectively, in the second, third and fourth columns of Table~\ref{fitresultsvslattice} for the $P_{1}^{2}(q^{2})$ (the other sequences yield almost identical results).
While the second and third columns lead similar results, the fourth column, including the updated FNAL/MILC form factor simulation of 2015, clearly shifts upwards(downwards) the $|V_{ub}|$$(f_{+}(0))$ value by about $1.3 \sigma$ yielding smaller statistical uncertainties and slightly enlarging the $\chi^{2}_{\rm{dof}}$.

Upon comparison with last column, we conclude that FNAL/MILC simulations of 2015 drives the form factor 
while disagreeing for more than $1\sigma$ with respect to all the other lattice simulations (including their own determination but from 2008). 
This fact explains why the 2016 PDG reported value has been shifted with respect to the earlier edition by $+1\sigma$.

\begin{table}[h!]
\centering
\begin{tabular}{llllll}
\hline
&HPQCD&HPQCD& HPQCD& All (Table~\ref{fitspectralattice})\\
&+RBC/UKQCD&+RBC/UKQCD&+RBC/UKQCD& \\
&&+MILC 2008&+MILC 2015& \\
\hline
$|V_{ub}|\times10^{3}$ &$3.34(12)(0)$&$3.29(12)(2)$&$3.54(9)(2)$&$3.53(8)(3)$ \\ 
$f_{+}(0)$ &$0.279(14)(1)$&$0.280(13)(4)$&$0.264(10)(1)$& 0.265(10)(2)\\
Pole (GeV) &5.35&5.30&5.29&5.29 \\
$\chi^{2}_{\rm{dof}}$ &1.52&1.27&1.43&1.27 \\
\hline                                           
\end{tabular}
\caption{Results for $|V_{ub}|$ and $f_{+}(0)$ as obtained with a $P_{1}^{2}(q^{2})$ approximant from simultaneous fits to $B\to\pi\ell\nu_{\ell}$ partial branching ratio experimental data \cite{delAmoSanchez:2010af,Lees:2012vv,Ha:2010rf,Sibidanov:2013rkk} and FF simulations obtained by different lattice Col.: HPQCD \cite{Dalgic:2006dt}, MILC 2008 \cite{Bailey:2008wp}, RBC/UKQCD \cite{Flynn:2015mha} and MILC 2015 \cite{Lattice:2015tia}.
Last column corresponds to our final results given in Table~\ref{fitspectralattice} collected here for ease of comparison. The first and second parenthesis refer, respectively, to the statistical and systematic uncertainties. The errors are symmetrized.}
\label{fitresultsvslattice}
\end{table}

We close this section by performing fits to the lattice FF predictions alone and extract $f_{+}(0)$.
This kind of exercise is new and, as a byproduct, allows us to determine $|V_{ub}|$ by equating the corresponding expression for the branching ratio $(\mathcal{BR})$ to the measured ones, $\mathcal{BR}(B^{0}\to\pi^{-}\ell^{+}\nu_{\ell})=(1.45\pm0.05)\times10^{-4}$ \cite{Olive:2016xmw}.
We only obtain reliable results when, at least, the HPQCD 2007 and MILC 2015 predictions are included into the data sets to be fitted and for approximants with two poles.
The corresponding fit results are gathered in Table~\ref{fitsLatticeonly}.

\begin{table}[h!]
  \centering
  \begin{tabular}{|c|l|l|c|c|c|c|c|c|c|c|}
\hline
    \multirow{1}{*}{Lattice data sets} & 
    \multicolumn{2}{c|}{\multirow{1}{*}{Approximant}} & 
    \multicolumn{1}{c|}{$\chi^{2}_{\rm{dof}}$}& 
    \multicolumn{1}{c|}{Poles (GeV)}&
    \multicolumn{1}{c|}{$f_{+}(0)$}&
    \multicolumn{1}{c|}{$|V_{ub}|\times10^{3}$}\\
    \cline{4-7}
    \hline
      \multirow{1}{*}{All} & \multirow{1}{*}{Free $B^{*}$} &$P_{2}^{0}$&0.57&$5.31, \,7.43$&$0.262(25)$&$3.66(6)$\\ \cline{3-7}
  \cite{Lattice:2015tia,Dalgic:2006dt,Bailey:2008wp,Flynn:2015mha}& \multirow{1}{*}{Fixed $B^{*}$} & \multirow{1}{*}{$P_{1,1}^{0}$} &$0.58$&$6.96$&$0.244(12)$&$3.80(7)$\\ \cline{3-7}
        \hline
    \hline
\multirow{1}{*}{\cite{Lattice:2015tia,Dalgic:2006dt} }  & \multirow{1}{*}{Free $B^{*}$} & \multirow{1}{*}{$P_{2}^{0}$} &$0.83$&$5.31, \, 7.42$&$0.260(25)$&$3.68(7)$\\ \cline{1-7}
    \hline          
  \end{tabular}
  \caption{Results for $|V_{ub}|$ and $f_{+}(0)$ obtained from fits to lattice FF simulations alone: HPQCD \cite{Dalgic:2006dt}, MILC 2008 \cite{Bailey:2008wp}, RBC/UKQCD \cite{Flynn:2015mha} and MILC 2015 \cite{Lattice:2015tia}.}\label{fitsLatticeonly}
\end{table}

\subsection{Unitary constraints on PA's fits}\label{unitaryconstraints}

All zeros and poles of our approximants must lie along the unitary branch cut in order to fulfill the unitary requirements that the FF imposes~\cite{Baker}. 
If a particular PA does not show this feature 
indicates the set of data fitted is not fulfilling the unitary requirements that must have. 
Thus, both defects and the appearance of poles and zeros outside the unitary branch cut are indications of a violation of unitary to a certain degree. 
Since we have performed a dedicated analysis Collaboration by Collaboration, bin by bin, and since we have found some cases which slightly violate these two statements, mostly happening when the BaBar 2012 data is involved, we are able to identify the source of the unitary deviation.


We find either complex-conjugate poles with an small imaginary part or a zero(s) within the radius of convergence for the $P_{2}^{0}(q^2)$ and $P_{1}^{1,2}(q^2)$ elements when fitting individually the BaBar 2012 data set, respectively. 
In particular, the complex-conjugate pole of $P_{2}^{0}(q^2)$ is found to be at $5.72\pm i0.53$ GeV while the zeros of the numerator are placed at $-4.88$ GeV and $-4.40$ GeV for the $P_{1}^{1}(q^2)$ and $P_{1}^{2}(q^2)$ elements, respectively (the second zero of $P_{1}^{2}(q^2)$ is placed at $12.97$ GeV, far away from the origin).
A complex-conjugate pole with an small imaginary part also shows up in the $P_{2}^{2}(q^2)$ element when performing the joint fit to data and lattice. 
In order to further explore on the origin of these extraneous poles and zeros we have also performed fits removing one experimental datum of each Collaboration e.g. those with more tension according to our Figures 2 and 3, and see what can we learn. 

In particular, we remove the fifth datum of BaBar 2011, the tenth of BaBar 2012 and Belle 2011 and the bin located at 9 GeV$^{2}$ of Belle 2013.
By doing this, we find that the zeros tend to move away from the radius of convergence while the complex-conjugate poles become cancelled by a close-by zero in the numerator, a \textit{Froissart} pole.

For example, a $P_{2}^{2}(q^2)$ approximant becomes effectively a $P_{1}^{1}(q^{2})$ after removing these four points and the fit to the rest of experimental data and lattice simulations yields $|V_{ub}|=3.57(8)\times10^{-3}$, $f_{+}(0)=0.260(9)$ with the pole located at $5.25$ GeV and $\chi^{2}_{\rm{dof}}=1.10$, while the same fit with all experimental points yielded, as collected in Table~\ref{fitspectralattice}, $|V_{ub}|=3.52(8)\times10^{-3}$, $f_{+}(0)=0.263(9)$ with a pole at $5.23$ GeV and $\chi^{2}_{\rm{dof}}=1.31$. The impact of these four points is remarkable, inducing a positive shift on the $|V_{ub}|$ after removing them by about $\Delta |V_{ub}| = 0.05\times10^{-3}$, a $0.4 \sigma$ deviation. 

Breaking of unitary is then reducing the value of $|V_{ub}|$ an enlarging the discrepancy between inclusive and exclusive determinations. 
In view of this fact and the difficulty on deciding the best strategy to take this unitary violation into account when dealing with experimental data (other strategies beyond removing bins could be envisaged), we have decided to add in quadrature the difference $\Delta |V_{ub}|$ as an extra source of error in our final determination of the CKM parameter and Eq.\,(\ref{centralresults1}) should be sound $|V_{ub}|=3.53(8)_{\rm{stat}}(6)_{\rm{syst}}\times10^{-3}$.
This error could be removed as soon as the experimental Collaborations could take our observation into account and explore systematically the potential unitary violation within their data sets.



\section{$B^{+}\to\eta^{(\prime)}\ell^{+}\nu_{\ell}$ decays and $\eta$-$\eta^{\prime}$ mixing}\label{section5}

In the previous section, the $B\to\pi$ form factor $f_{+}(q^{2})$ has been parameterized using PAs to fit experimental data on the $B\to\pi\ell\nu_{\ell}$ differential branching ratio distribution w/o lattice FF simulations. 
In this section, we would like to take advantage of these parameterizations to describe the $B^{+}\to\eta^{(\prime)}\ell^{+}\nu_{\ell}$ decays as discussed in the following.

The expression for the differential $B^{+}\to\eta^{(\prime)}\ell^{+}\nu_{\ell}$ decay width is given by the same expression as for the $B\to\pi\ell\nu_{\ell}$ decay mode in Eq.~(\ref{decayrate}) by replacing the final state pion by $\eta^{(\prime)}$
\bea
\frac{d\Gamma(B^{+}\to\eta^{(\prime)}\ell^{+}\nu_{\ell})}{dq^{2}}=\frac{G_{F}^{2}|V_{ub}|^{2}}{192\pi^{3}m_{B}^{3}}|p_{\eta^{(\prime)}}|^{3}|f_{+}^{B^{+}\eta^{(\prime)}}(q^{2})|^{2}\,,
\label{decayrateEta}
\eea
where now $f_{+}^{B^{+}\eta^{(\prime)}}(q^{2})$ represents the hadronic $B^{+}\to\eta^{(\prime)}$ transition. 
What the $B^{+}\to\eta^{(\prime)}$ transition is probing is the light-quark content of the $\eta^{(\prime)}$ mesons since the $s\bar{s}$ component can only be accessed via a $B_s$ meson decay. 
This is so because from the quark-flavour perspective, $\eta^{(\prime)}$ mesons are an admixture of  $u\bar{u}, d\bar{d}$ and $s\bar{s}$ components. 
Defining $|\eta_{q}\rangle=\frac{1}{\sqrt{2}}|u\bar{u}+d\bar{d}\rangle$ and $|\eta_{s}\rangle=|s\bar{s}\rangle$ in this quark-flavour basis, one can relate the mathematical $|\eta_{q,s} \rangle$ states with the physical $|\eta^{(\prime)} \rangle$ ones through the following matrix rotation
\bea
\left(
\begin{array}{cc}
\eta \\[0.5ex]
\eta^{\prime}
\end{array}
\right)
=
\left(
\begin{array}{cc}
\cos\phi & -\sin\phi\\[0.5ex]
\sin\phi & \cos\phi
\end{array}
\right)\left(
\begin{array}{cc}
\eta_{q}\\[0.5ex]
\eta_{s}
\end{array}
\right)\,,
\eea
where $\phi$ gives the degree of admixture.


Contrary to $f^{B\pi}_{+}(q^{2})$, there are no, to the best of our knowledge, FF simulations of $f_{+}^{B^{+}\eta^{(\prime)}}(q^{2})$ on the lattice while only a few calculations at $q^{2}=0$ exist \cite{Yao:2018tqn,Ball:2004ye,Ball:2007hb}. 
Therefore, we will relate the $f_{+}^{B^{+}\eta^{(\prime)}}(q^{2})$ with the $f^{B\pi}_{+}(q^{2})$  using the quark-flavor basis. 
Assuming isospin symmetry between the $u$ and $d$ quarks, the form factor $f_{+}^{B^{+}\eta^{(\prime)}}(q^{2})$ can be related to the $f_{+}^{B^{+}\pi^{0}}(q^{2})$ ones through \cite{Kim:2001xi}
\begin{eqnarray}
\nonumber&&f_{+}^{B^{+}\eta}(q^{2}) = \cos\phi f_{+}^{B^{+}\eta_{u\bar{u}}}(q^{2})\simeq\cos\phi f_{+}^{B^{+}\pi^{0}}(q^{2})\,,\\ 
&&f_{+}^{B^{+}\eta^{\prime}}(q^{2}) = \sin\phi f_{+}^{B^{+}\eta_{u\bar{u}}}(q^{2})\simeq\sin\phi f_{+}^{B^{+}\pi^{0}}(q^{2})\,.
\label{BtoetasFF}
\end{eqnarray}
taking $\eta_{u\bar{u}}\simeq\pi^0$ as in Refs.~\cite{Escribano:2015nra,Masjuan:2017tvw}.

\begin{figure}[h!]
\begin{center}
\includegraphics[scale=0.7]{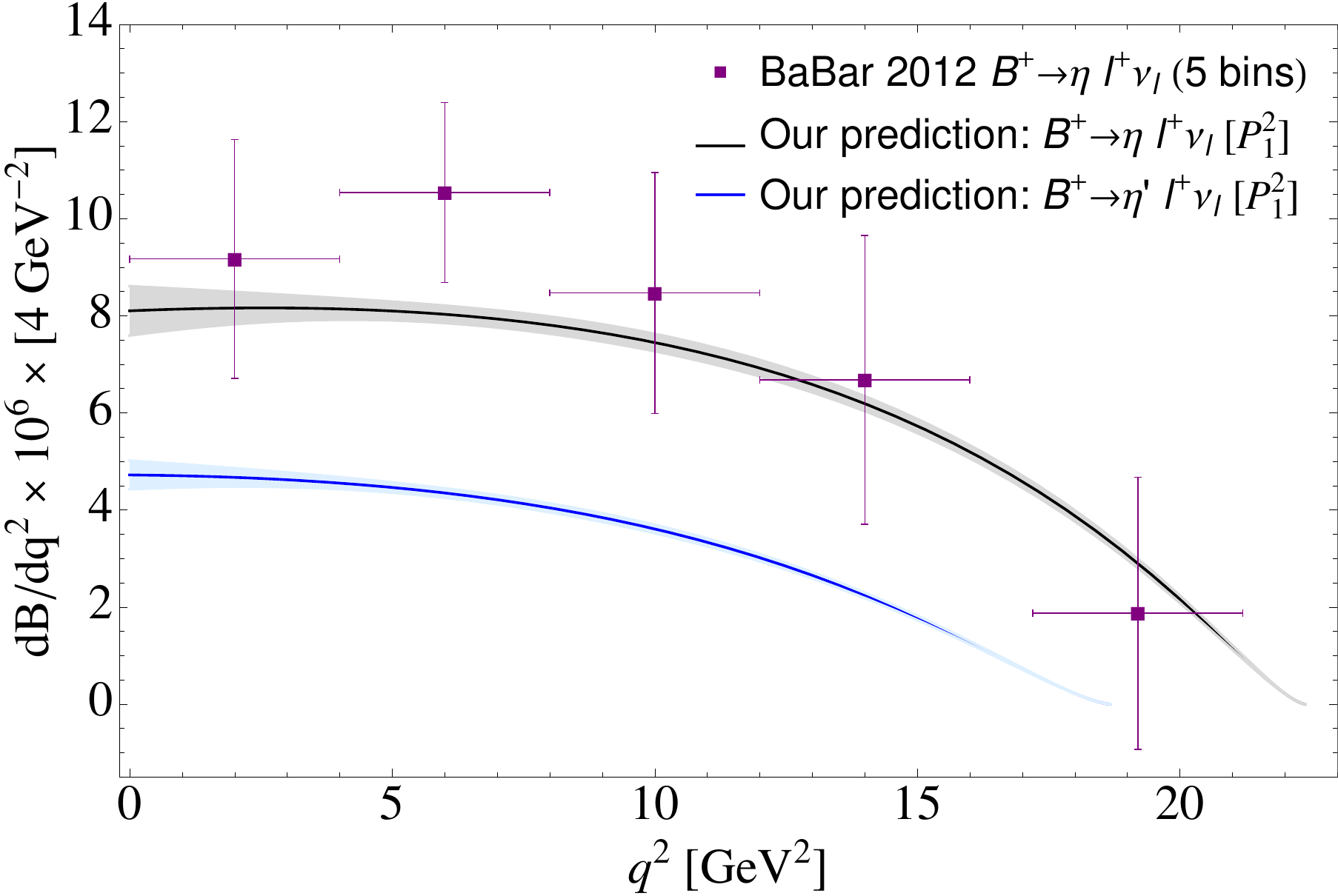}
\caption{\label{Btoetaspectrapred}Predictions for the $B^{+}\to\eta\ell^{+}\nu_{\ell}$ (black) and $B^{+}\to\eta^{\prime}\ell^{+}\nu_{\ell}$ (blue) differential branching ratio distribution. 
The BaBar 2012 experimental data is taken from \cite{Lees:2012vv}.}
\end{center}
\end{figure}

From Eqs.~(\ref{decayrateEta}) and (\ref{BtoetasFF}), and taking $f^{B\pi}_{+}(q^{2})$ from any of the descriptions given in Table~\ref{fitspectralattice} together with the corresponding values for $|V_{ub}|$, we can describe the differential branching ratio distribution of the $B^{+}\to\eta^{(\prime)}\ell^{+}\nu_{\ell}$ decays by setting the numerical value of the $\eta$-$\eta^{\prime}$ mixing angle to, for example, $\phi=38.3(1.6)$ \cite{Escribano:2015nra}.
Our prediction for the $B^{+}\to\eta\ell^{+}\nu_{\ell}$ differential branching fraction distribution is shown and compared with BaBar 2012 measurements in 5 bins of $q^{2}$ in Fig.~\ref{Btoetaspectrapred} for $P_{1}^{2}(q^{2})$ (black solid curve). 
Our description is seen quite in accordance with data although the second experimental datum seems to be slightly in tension. 
In Fig.~\ref{Btoetaspectrapred} we also show our prediction for the $B^{+}\to\eta^{\prime}\ell^{+}\nu_{\ell}$ branching ratio distribution (blue solid curve), in this case without any experimental data to compare with. 

One interesting information we can extract from the curves displayed in Fig.~\ref{Btoetaspectrapred} is the corresponding integrated branching ratios. 
We obtain
\bea
\mathcal{BR}(B^{+}\to\eta\ell^{+}\nu_{\ell})=0.34(2)(1)\times10^{-4}\,,\,\mathcal{BR}(B^{+}\to\eta^{\prime}\ell^{+}\nu_{\ell})=0.15(1)(1)\times10^{-4}\,,
\label{brpredictions}
\eea
where the first and second error correspond, respectively, to the uncertainty associated to $|V_{ub}|$ (cf.~Table~\ref{fitspectralattice}) and to the form factor.
These values are $0.5\sigma$ and $1\sigma$ away from the branching ratio measurements $\mathcal{BR}(B^{+}\to\eta\ell^{+}\nu_{\ell})=0.38(5)(5)\times10^{-4}$ and $\mathcal{BR}(B^{+}\to\eta^{\prime}\ell^{+}\nu_{\ell})=0.24(8)(3)\times10^{-4}$ reported by BaBar in 2012 \cite{Lees:2012vv}, respectively.
In light of these results, we conclude that the simple mixing scheme assumed in Eq.~(\ref{BtoetasFF}) works quite well for $B^{+}\to\eta\ell^{+}\nu_{\ell}$ decay while for $B^{+}\to\eta^{\prime}\ell^{+}\nu_{\ell}$ is not conclusive due to the large uncertainty on the experimental measurement.
In order to go beyond the simple quark-flavour basis decomposition, we would like to encourage the experimental groups to measure again these channels with improved precision. 

We can also perform the exercise of letting the mixing angle $\phi$ to float and determine its value by equating the branching ratios as obtained in Eq.~(\ref{brpredictions}) to their corresponding experimental values. 
In the spirit of Ref.\,\cite{DiDonato:2011kr}, we equate ratios of branching ratios in order to eliminate the uncertainties associated to $|V_{ub}|$. 
Similarly, we could also compare with the $B\to\pi\ell\nu$ BaBar measurement, $\mathcal{BR}(B^{+}\to\pi^{0}\ell^{+}\nu)=0.77(4)(4)\times10^{-4}$~\cite{Lees:2012vv}.
The corresponding ratios of branching ratios, after symmetrizing errors, read
\begin{eqnarray}
\nonumber&&R_{\eta^{\prime}/\eta}\equiv\frac{\mathcal{BR}(B\to\eta^{\prime}\ell\nu_{\ell})}{\mathcal{BR}(B\to\eta\ell\nu_{\ell})}=0.63(25)\,,\\
\nonumber&&R_{\eta^{\prime}/\pi}\equiv\frac{\mathcal{BR}(B\to\eta\ell\nu_{\ell})}{\mathcal{BR}(B\to\pi\ell\nu_{\ell})}=0.49(10)\,,\\
&&R_{\eta/\pi}\equiv\frac{\mathcal{BR}(B\to\eta^{\prime}\ell\nu_{\ell})}{\mathcal{BR}(B\to\pi\ell\nu_{\ell})}=0.31(11)\,,
\label{ratioBR}
\end{eqnarray}
that we equate to (cf. Eq.~(\ref{BtoetasFF}))
\begin{eqnarray}
\nonumber R_{\eta^{\prime}/\eta}&=&|\tan\phi|^{2}\frac{\int_{0}^{(m_{B}-m_{\eta^{\prime}})^{2}}dq^{2}|p_{\eta^{\prime}}|^{3}|F_{+}^{B^{+}\pi^{0}}(q^{2})|^{2}}{\int_{0}^{(m_{B}-m_{\eta})^{2}}dq^{2}|p_{\eta}|^{3}|F_{+}^{B^{+}\pi^{0}}(q^{2})|^{2}}\,,\\
\nonumber R_{\eta^{\prime}/\pi}&=&|\sin\phi|^{2}\frac{\int_{0}^{(m_{B}-m_{\eta^{\prime}})^{2}}dq^{2}|p_{\eta^{\prime}}|^{3}|F_{+}^{B^{+}\pi^{0}}(q^{2})|^{2}}{\int_{0}^{(m_{B}-m_{\pi})^{2}}dq^{2}|p_{\pi}|^{3}|F_{+}^{B^{+}\pi^{0}}(q^{2})|^{2}}\,,\\
R_{\eta/\pi}&=&|\cos\phi|^{2}\frac{\int_{0}^{(m_{B}-m_{\eta})^{2}}dq^{2}|p_{\eta}|^{3}|F_{+}^{B^{+}\pi^{0}}(q^{2})|^{2}}{\int_{0}^{(m_{B}-m_{\pi})^{2}}dq^{2}|p_{\pi}|^{3}|F_{+}^{B^{+}\pi^{0}}(q^{2})|^{2}}\,.
\label{BRexpressions}
\end{eqnarray}

The corresponding results are collected in Table~\ref{mixingprediction}. 
We observe that the central values show some scatter though they all agree within errors due to the large uncertainties in Eq.~(\ref{ratioBR}).
In order to extract the mixing angle $\phi$ from $B\to\eta^{(\prime)}$ transitions with more precision, measurements of these decays with higher precision are required.

\begin{table}[h!]
\centering
\begin{tabular}{|c|c|c|c|c|c|}
\hline
Mixing angle& $R_{\eta^{\prime}/\eta}$ &$R_{\eta/\pi}$ &  $R_{\eta^{\prime}/\pi}$\\
\hline
$\phi\,(^{\circ})$ & $43.3\pm6.0$ & $37.6\pm8.0$ & $48.2\pm11.0$ \\
\hline
\end{tabular}
\caption{Predictions for the $\eta$-$\eta^{\prime}$ mixing $(\phi)$ as obtained by equating Eq.~(\ref{ratioBR}) to Eq.~(\ref{BRexpressions}).}
\label{mixingprediction}
\end{table}

As a final exercise, we would also liked to fit either the individual BaBar 2012 $B^{+}\to\eta\ell^{+}\nu_{\ell}$ decay experimental data or perform a combined fit to $B^{+}\to\pi\ell^{+}\nu_{\ell}$ and $B^{+}\to\eta\ell^{+}\nu_{\ell}$ decays with the goal to provide an alternative semileptonic charmless $B$ decay determination of $|V_{ub}|$. 
However, due to the poor experimental situation in the case of $B^{+}\to\eta\ell^{+}\nu_{\ell}$, we decide to postpone this analysis for the future.

\section{Conclusions}\label{conclusions}

In this paper we have reexamined the $B\to\pi\ell\nu_{\ell}$ decays to extract the CKM parameter $|V_{ub}|$ based on experimental data, lattice calculations and unitary constraints of the participant form factor. 
Contrary to the most commonly used $z$-expansion and Vector Meson Dominance models, we perform our analysis based on the method of Pad\'{e} Approximants after realizing that most of the recent previous analyses belong to the Pad\'{e} Theory, even though no one mention it. 
Thus, the rules and constrains imposed by the convergence theorems for Pad\'{e} Approximants to the form factor, so far neglected, are fully exploited here, allowing to ascribe to our final result a new source of systematic or truncation error.

From our dedicated analysis we obtain $|V_{ub}|=3.53(8)_{\rm{stat}}(6)_{\rm{syst}}\times10^{-3}$.
This quantity includes both statistical, from the fitted data, and systematic, from the truncation of the Pad\'{e} sequence, uncertainties, and has been obtained guaranteeing the independency with respect of the model calculation of $f_{+}(0)$ as external constrain.

On a first stage, after a detailed review of the state-of-the-art experimental data, determinations of $|V_{ub}|$ and theoretical representations of the analytical structure of the form factor, we have analyzed the measured $q^{2}$ differential branching ratio distribution experimental data released by the BaBar and Belle Collaborations.
Our fitting strategy started by performing a combined analysis to all data sets using different types of Pad\'{e} sequences.
We thus have determined first the product $|V_{ub}f_{+}(0)|$ directly from the fits and then extracted the CKM element $|V_{ub}|$ by invoking external theoretical information on $f_{+}(0)$.
The resulting fit results are presented in Table \ref{fitspectra} and a graphical account provided in Fig.\,\ref{spectrum}.
We then have carried out a detailed analysis Collaboration by Collaboration.
The outcome of the individual fits is displayed in Fig.\,\ref{distcoll} and the neat effect on each experimental datum due to fitting all experimental data together with respect to fitting data of each Collaboration separately is shown in Fig.\,\ref{sigmapoints}.
This exercise allow us to classify the four differential experimental data sets according to their increasing degree of robustness: BaBar 2011, Belle 2013, Belle 2011 and BaBar 2012.

On a second stage, we have included into the analysis the four available lattice QCD predictions on the form factor shape.
This data dominates the large-$q^{2}$ region and it is essential for a precise determination of $|V_{ub}|$. 
The corresponding fit results are collected in Table \ref{fitspectralattice} indicating that the statistical uncertainty associated to $|V_{ub}|$ is reduced by $\sim80\%$ after the inclusion of lattice data.
We have also found that, out of the four lattice form factor simulations, the predictions released by the MILC Collaboration in 2015 tends to drive the form factor (see Table \ref{fitresultsvslattice}) but slightly enlarging the $\chi^{2}_{\rm{dof}}$. 
As a byproduct of our analysis, we have predicted the BCL form factor series coefficients that are obtained by matching the corresponding Taylor series expansion.
The coefficients thus obtained are shown and compared with the determinations given by lattice groups in Table \ref{fittedcoefficients} while the $q^{2}$ shape of the reconstructed BCL parameterization is displayed in Fig.\,\ref{FF_joint_fit} proving the ability of the Pad\'{e} Approximants in this transition.

On a third stage, motivated by the impact of the lattice data, we have also explored fits to the lattice predictions alone.
The fit results are shown in Table \ref{fitsLatticeonly} reflecting that only those approximants with two poles have the ability to extract first $f_{+}(0)$ and then determine $|V_{ub}|$ by equating the theoretical expression for the branching ratio to the corresponding experimental measurement.

\begin{figure}[h!]
\centering\includegraphics[scale=0.7]{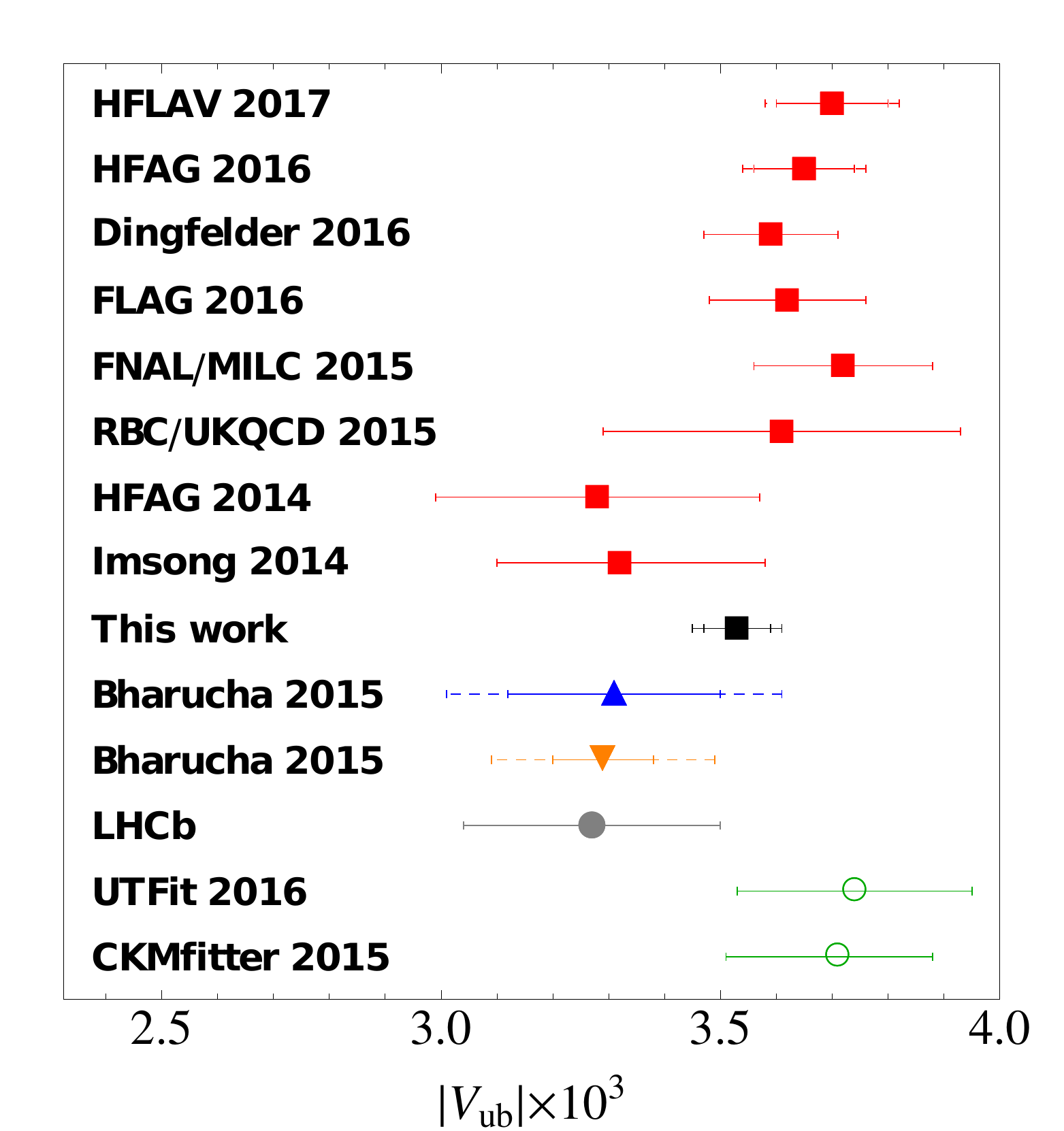}
\caption{Status of $|V_{ub}|$ determinations from the exclusive $B\to\pi\ell\nu_{\ell}$ decays $(\textcolor{red}{\blacksquare})$ including HFLAV 2017 \cite{Amhis:2016xyh}, HFAG 2016~\cite{HFAG2016}, Dingfelder 2016~\cite{Dingfelder:2016twb}, FLAG 2016~\cite{Aoki:2016frl}, FNAL/MILC 2015~\cite{Lattice:2015tia}, RBC/UKQCD 2015~\cite{Flynn:2015mha}, HFAG 2014~\cite{Vubpdg2014}, Imsong 2014~\cite{Imsong:2014oqa} and this work $(\textcolor{black}{\blacksquare})$, from $B\to\omega\ell\nu_{\ell}$ $(\textcolor{blue}{\blacktriangle})$ and $B\to\rho\ell\nu_{\ell}$ $(\textcolor{orange}{\blacktriangledown})$ Bharucha 2015~\cite{Straub:2015ica} and from $\Lambda_{b}\to p\mu\nu_{\mu}$ $(\textcolor{gray}{\bullet})$ LHCb~\cite{Detmold:2015aaa}, and from indirect fits $({\color{green!65!blue}{\circ}})$ UTFit 2016~\cite{Bona} and CKMfitter 2015~\cite{Charles:2004jd}. 
The solid and dashed error bar account, respectively, for the statistical and systematic uncertainties.}
\label{Vuboutlook}
\end{figure}

Our central result, $|V_{ub}|=3.53(8)_{\rm{stat}}(6)_{\rm{syst}}\times10^{-3}$, is presented and compared with other determinations using other methods and fitted data sets in Fig.~\ref{Vuboutlook}.
We would like to remark two features concerning this value that are related to the use of Pad\'{e} Approximants. 
The first one, is that the central value tends to fall slightly downwards with respect to the values determined with the $z$-expansion parameterization in the studies carried out in the recent years.
And the second one, is that the method allow us to ascribe a systematic uncertainty from the truncated Pad\'{e} sequence.
In fact, the $z$-paramaterizations do also allow to attribute a systematic error following the same reasoning.
However, in practice, it has not so far usually been considered.
For example, based on our criterion, the result as obtained by the FNAL/MILC Collaboration in 2015 would read $|V_{ub}|=3.72(16)_{\rm{stat}}(9)_{\rm{syst}}$, where the systematic uncertainty stems from the differing results for $N=3,4$ (cf. Eq.\,(\ref{zparam2})).  
In our study, the ascribed systematic uncertainty includes, for the first time, an additional conservative source of error due to the unitarity constraints discussed in section \ref{unitaryconstraints}.
These constraints have to do with the appearance of extraneous poles and zeros outside the unitary branch cut and might indicate, to a certain degree, violations of unitarity.

As a final concluding remark for the $B\to\pi\ell\nu_{\ell}$ decays, we would like to point out that, contrary to the $z$-expansion and VMD models where the $B^{*}$ pole position is fixed to $5.325$ GeV in advance, a very competitive value for $|V_{ub}|$ can be extracted without imposing any information regarding the position of it as we have shown along the lines of our detailed analysis.

In the second part of this work, we have addressed the $B^{+}\to\eta^{(\prime)}\ell^{+}\nu_{\ell}$ decays taking advantage of the $B\to\pi$ form factor parameterizations derived in the first part.
In particular, we relate the participant $B\eta^{(\prime)}$ form factor to the $B\pi$ ones by a single Euler angle rotation assuming that the light-quark component of the $\eta^{(\prime)}$ is a $q\bar{q}$ pion to a large extent.
Under this simple assumption, we obtain a reliable prediction for the differential branching ratio distribution of the $B^{+}\to\eta\ell^{+}\nu_{\ell}$ decay as shown in Fig.\,\ref{Btoetaspectrapred} compared to the BaBar measurement in 5 bin of $q^{2}$ released in 2012.  
As a byproduct of our study, we have also extracted the $\eta$-$\eta^{\prime}$ mixing angle. This quantity, however, carries a large statistical error due to the large uncertainty on the measured $B^{+}\to\eta^{(\prime)}\ell^{+}\nu_{\ell}$ branching ratios.
Regarding our prediction for the $B^{+}\to\eta^{\prime}\ell^{+}\nu_{\ell}$ decay distribution, there is no experimental data to compare with so far.
In order to go beyond the simple quark-flavour basis decomposition and extract the $\eta$-$\eta^{\prime}$ mixing angle with improved precision we would like to encourage experimental groups to measure these semileptonic $B^{+}\to\eta^{(\prime)}$ transitions with improved precision.

\newpage

\section*{Acknowledgements}

The authors acknowledge Yuzhi Liu for discussions.
The work of S.G-S is supported in part by the CAS President's International Fellowship Initiative for Young International Scientists (Grant No. 2017PM0031), by the Sino-German Collaborative Research Center \textquotedblleft Symmetries and the Emergence of Structure in QCD\textquotedblright\,(NSFC Grant No. 11621131001, DFG Grant No. TRR110), and by NSFC (Grant No. 11747601). The work of P.M by the Beatriu de Pin\'os postdoctoral programme of the Government of Catalonia's Secretariat for Universities and Research of the Ministry of Economy and Knowledge of Spain.

\end{document}